\documentclass[fleqn,usenatbib,useAMS]{mnras}

\usepackage{graphicx}	
\usepackage{amsmath}	
\usepackage{amssymb}	
\usepackage{multicol}        
\usepackage{bm}		
\usepackage{pdflscape}	
\DeclareGraphicsExtensions{.pdf,.png,.jpg,.jpeg,.eps}
\graphicspath{{./Figures/}}
\usepackage{acronym}
\usepackage{subcaption}
\usepackage{xcolor}
\usepackage[most]{tcolorbox}

\newcommand{\avgB}{\langle B \rangle}
\newcommand{\avgrho}{\langle \rho \rangle}
\newcommand{\resunit}{cm$^2$s$^{-1}$}

\usepackage[T1]{fontenc}
\usepackage{ae,aecompl}

\usepackage{txfonts}
\usepackage{float}
\usepackage{caption}
\usepackage{lineno}

\title[Magnetic fields in NSs from MHD simulations]{Magnetic field configurations in neutron stars from MHD simulations}

\author[Ankan Sur]{Ankan Sur$^{1}$\thanks{Contact e-mail: \href{mailto:ankansur@camk.edu.pl}{ankansur@camk.edu.pl}}, Brynmor Haskell$^{1}$, Emily Kuhn$^2$
	\\
	$^{1}$Nicolaus Copernicus Astronomical Center, Polish Academy of Sciences, Bartycka 18, 00-716 Warsaw, Poland\\
	$^{2}$Department of Physics, Yale University, New Haven, CT 06520, United States\\
    }
\date{Last updated \today; in original form \today}

\begin{document}
	\maketitle
	
\begin{abstract}
	We have studied numerically the evolution of magnetic fields in barotropic neutron stars, by performing nonlinear magnetohydrodynamical simulations with the code PLUTO. For both initially predominantly poloidal and toroidal fields, with varying strengths, we find that the field settles down to a mixed poloidal-toroidal configuration, where the toroidal component contributes between $10\%$ and $20 \%$ of the total magnetic energy.
This is, however, not a strict equilibrium, as the instability leads to the development of turbulence, which in turn gives rise to an inverse helicity cascade, which determines the final ``twisted torus'' setup. The final field configuration is thus dictated by the non-linear saturation of the instability, and is not stationary. The average energy of the poloidal and toroidal components, however, is approximately stable in our simulations, and a complex multipolar structure emerges at the surface, while the magnetic field is dipolar at the exterior boundary, outside the star.


\end{abstract}


\newcommand{\gpercc}{\ensuremath{\mbox{gm cm}^{-3}}}
\newcommand{\rstar}{\ensuremath{\mbox{$R_{\star}$}}}
\newcommand{\alfven}{\ensuremath{\mbox{Alfv\'{e}n}}}

\section{Introduction\label{sec:introduction}}
Neutron star(s) (NS(s)) are extremely dense compact objects bearing the strongest magnetic fields known to date in the universe. The surface field strength for ordinary NSs ranges from $10^{12}-10^{13}$G while for magnetars, it goes well beyond $10^{15}$G. Despite such estimates for the strength of the magnetic field, its structure is not completely known to us. Polarimetric studies of radio emission from pulsars have been used to probe the geometry of pulsar magnetospheres. Such observations favour a predominantly dipolar magnetic field, although there is evidence for higher multipoles \citep{Chung1, Chung2, Jonas2020}.
Recent observations in X-rays by the NICER mission have confirmed that the field at the surface is far from an aligned dipole, but rather an intricate multipolar structure is present \citep{Nicer1}.

The internal field topology is even more difficult to probe directly with observations, but is thought to play a fundamental role in determining the nature and strength of electromagnetic and gravitational wave emission of the star \citep{Thompson96, Cutler2002, Guver2011}. As the field of gravitational wave astronomy advances, it may, in fact, be possible to use gravitational-wave signatures to discriminate between different magnetic field topologies \citep{Lasky2013}. It is thus of great importance to obtain a theoretical understanding of the interior field and use numerical simulations to explore its nature.

Several studies have been carried out to investigate the equilibrium configuration of magnetic main sequence stars and white dwarfs, in which the Lorentz force is balanced by pressure and gravity; e.g. for the axisymmetric case by \cite{Braithwaite:2005md, Braithwaite:2005xi, Braithwaite:2007fy, Armaza2015} and a similar non-axisymmetric study by \cite{Braithwaite2008}. For the NS case, equilibrium solutions in Newtonian gravity were obtained by \cite{Haskell:2007bh, lander2009magnetic, lander2010oscillations, lander2011oscillations, Herbrik2017, Frederick:2020yhw} and in general relativity by \cite{Kiuchi2008, Ciolfi:2010td,Ciolfi2013, Pili14, Pili17} and the role of stratification was investigated by \citet{Kostas2012, Reisenegger2009}.
Finding equilibrium and stability conditions for the magnetic field in stars has been an important long-standing question that dates back to earlier studies by \cite{Chandra53, Tayler57, Tayler73, Wright73, Markey73, Markey74, Flowers77}. A purely poloidal field undergoes the so-called ``Taylor instability'' and is thus unstable \citep{Ferraro54, Monaghan, Bocquet95}. In NSs this instability has been studied numerically in general relativity by \citet{Ciolfi:2011xa, Lasky:2011un, Ciolfi2012}, who confirm that an initially poloidal field is unstable on the order of an $\alfven$ crossing timescale and toroidal components of the field are generated. The equilibrium configuration is often approximated as a \textit{twisted-torus} configuration where a toroidal component stabilises the poloidal field \citep{Braithwaite:2005xi, Braithwaite:2007fy} or a \textit{tilted-torus}  configuration with the magnetic axis tilted with respect to the rotation axis \citep{Lasky2013}, as the inclination angle between the two grows \citep{Lander2018, Lander2019}. An initial purely toroidal field is also unstable to the azimuthal wavenumber m=1 mode of oscillation which is independent of the field strength but instead depends on the geometry \citep{Roxburgh66, Tayler73} and can lead to strong poloidal components developing.

Several open questions remain, however, as to the exact configuration the field will settle down to. Specifically, while it is clear that a mixed field is required for stability, the relative strength of the components cannot be obtained directly from the study of equilibrium configurations which generally allow a degree of freedom in tuning this parameter (see \citealt{GlampLasky2016} for an in depth discussion of this issue). It is thus possible to obtain models in which the toroidal field strength ranges from a few percent of \citep{lander2009magnetic} to more than an order of magnitude higher \citep{Ciolfi2013} than the strength of the poloidal component.

In fact, the stability of barotropic stars has been questioned by \cite{lander2012,Mitchell2015} who hypothesize that all barotropic models are unstable, while \citet{Reisenegger2009, Akgun2013} suggest that stratification plays an important role in stabilising the field, and non-barotropic models of magnetised stars are stable.
Since most hydromagnetic instability studies have focused on building equilibrium configurations starting with a specific choice of geometry, fully non-linear time evolution for NSs for a range of initial topologies and a barotropic EOS need to be carried out, to determine not only whether the field is unstable, but also, crucially, what the final state determined by the non-linear saturation of the instability is. It is also important to understand how magnetic helicity is generated and transferred in the star as the instability proceeds. Both superfluid and standard MHD turbulence are expected in NS interiors, and the evolution of the field, especially soon after birth when the star is still differentially rotating, is likely linked to the action of a dynamo in the interior \citep{Thompson1993}.

It is crucial to obtain an understanding of these issues, as the field configuration of a NS plays an important role in attempts to determine the mass and radius of the star from X-ray observations \citep{Nicer2, Nicer3}, and in determining the gravitational wave emission properties of the system \citep{LaskyRev}.

To address the problem in this paper we perform non-linear magnetohydrodynamical (MHD) simulations of magnetised NSs to characterize the instability, the global evolution, and the final configuration of the magnetic field. We neglect effects due to superfluidity and superconductivity in the core and do not model the crust of the star. These choice are partly due to numerical convenience, but also reflect the fact that we are modelling instabilities on dynamical time scales, which will determine the configuration of the field shortly after the NS is born.
 After birth, the star cools down and there is a window during the first few hours of life, where differential rotation has likely been dissipated, but the crust has not yet formed and matter is not yet superfluid, thus justifying an ideal MHD description. Our simulation is relevant in such a scenario, as one would expect this setup to be `frozen' in \citep{Ciolfi:2010td}. This field configuration to which the star settles can thus be used as initial conditions for evolution on longer timescales of $10^3-10^5$ years, over which the Hall effect, ambipolar diffusion and Ohmic dissipation will affect the magnetic field \citep{GR92,PG07}. We work in Newtonian gravity as general relativity generally does not affect the qualitative nature of the magnetic instabilities \citep{Siegel2013}, and this allows us to explore a larger portion of parameter space.
We explore different setups, both in resistive and ideal MHD, and different initial conditions, which allow for fields with initially stronger poloidal or toroidal components. The initial field generally goes unstable on an $\alfven$ crossing timescale and we follow the development of the instability, which leads to the development of turbulence in the system, which in turn seeds the growth of magnetic helicity.

Our results show that in general the system reaches turbulent equilibrium, in which the average field strengths settle down to a stable ratio. In all our final field configurations, including those with initially stronger toroidal fields, the field is predominantly poloidal, but a weaker toroidal component ($\lesssim 20\%$ of the total magnetic energy) is present.

 The article is arranged as follows: In section 2, we discuss our numerical setup; In section 3, we discuss our results for the different setups considered in our simulations; section 4 discusses the effect of resistivity and section 5 the onset of turbulence. The convergence of our results is discussed in section 6, and finally conclusions and discussions are presented in section 7.

\vspace{0.4 cm}
\section{Physical system and Numerical Setup\label{sec:setup}}
We use the publicly available code PLUTO \footnote{http://plutocode.ph.unito.it/} by \cite{Mignone:2007iw} to solve the MHD equations [\ref{mhd1}-\ref{mhd4}].

\begin{eqnarray}
\centering
\frac{\partial \rho}{\partial t} + \nabla \cdot (\rho \bm{v}) = 0
\label{mhd1}
\end{eqnarray}
\begin{eqnarray}
\frac{\partial \bm{v}}{\partial t} + \bm{v} \cdot \nabla  \bm{v} + \frac{1}{4\pi\rho}\bm{B}\times (\nabla \times \bm{B}) + \frac{1}{\rho}\nabla p = -\nabla \Phi 
\end{eqnarray}
\begin{eqnarray}
\frac{\partial \bm{B}}{\partial t} + \bm{B}(\nabla \cdot \bm{v}) - (\bm{B} \cdot \nabla)\bm{v} + (\bm{v} \cdot \nabla)\bm{B} = 0
\label{mhd3} 
\end{eqnarray}

\begin{eqnarray}
\frac{\partial p}{\partial t}+ \bm{v} \cdot \nabla p + \rho c_s^2\nabla \bm{v} = 0,
\label{mhd4}
\end{eqnarray}
where $c_s$ is the sound speed. The above set of equations are closed with a barotropic EOS given by $p=p(\rho)$, which we take to be an $n=1$ polytrope. Although, the initial parameters are defined in terms of primitive variables ($p, \rho, \bm{v}, \bm{B}$), computations are done using conservative variables ($\rho, \rho\bm{v}, E, \bm{B} $), where $E = \rho\epsilon + \rho \bm{v}^2/2 + \bm{B}^2/2$. The above set of equations are solved, except for equation \ref{mhd4} where the pressure is calculated using the EOS and the density (helping to maintain the barotropy of the system), in a spherical coordinate system in 3 dimensions using a static grid which is divided into a number of points with $N_r$ in the radial direction $r$, $N_{\theta}$ in the polar direction $\theta$, and $N_{\phi}$ in the azimuthal direction $\phi$. However, our $r$-grid is non-uniform having a resolution ($\Delta r \sim$ 0.19 km) inside the star as compared to the atmosphere (where $\Delta r \sim$ 0.25 km). Interpolations are done with a piece-wise parabolic function which is accurate to second order in space. A Runge Kutta 3 (RK3) time stepping is used and we set the Courant-Friedrichs-Lewy (CFL) limit to 0.3. We use a Harten-Lax-van Leer (HLL) Riemann solver for computing the fluxes. The solenoidal constraint $\nabla\cdot\bm{B} = 0$ is maintained using the hyperbolic divergence cleaning method. The code does not solve the Poisson equation. We analytically solve for the gravitational potential in different regions of the star and provide it as an input. Our gravitational field does therefore not evolve with time. The density distribution of the star is, however, only very weakly affected by the magnetic field and this is generally a good approximation \citep{Haskell:2007bh}.

\subsection{{Initial Conditions}} \label{ICs}

We consider a non-rotating star which is modeled by solving the Lane-Emden equation with $n=1$ polytrope such that \footnote{PLUTO does not have an inbuilt barotropic EOS. We have suitably modified the ISOTHERMAL EOS such that the proportionality constant $k_{\star}$ remains fixed.} $P = k_{\star}\rho^2$, where $k_{\star}=4.25 \times 10^4$ {gm$^{-1}$cm$^{-5}$s$^{-2}$} \footnote{All our work is carried out using CGS units.}. We choose a background star with a total mass of 1.4$M_{\odot}$, radius $R_{\star}=10$ km and central density $\rho_c = 2.17 \times 10^{15}$ \gpercc. The density of the star only has radial dependence given by,

\begin{linenomath*}
\begin{equation}
\rho = \rho_c \frac{\sin y}{y}
\label{rhoeq}
\end{equation}
\end{linenomath*}
where $y=\frac{\pi r}{R_{\star}}$. As in most numerical MHD studies, it is necessary to replace the vacuum outside the stellar surface with an atmosphere of low-density fluid, in order to avoid computational difficulties due to diverging $\alfven$ velocities as the density falls to zero. Equation \ref{rhoeq} shows that $\rho$ falls rapidly and vanishes while approaching the edge of the star. Since the atmosphere has a non-zero density $\rho_{\text{atm}}$, this would cause a sharp gradient across the boundary of the star. In order to prevent such unrealistic jumps in density at the surface, we cut the star at a radius of $r=0.975 \rstar$  so that $\rho(r<=R) > \rho_{\text{atm}}$.{ We set $\rho_{atm} = 10^{12}$ {\gpercc} and explore two different setups, one in which the atmosphere extends up to a distance of 1.2 {\rstar} with no resistivity while another which extends up to 2{\rstar} and includes a resistive layer in the atmosphere of the star, which we will discuss in detail in the following sections.}
We start our simulation with two different initial conditions. The first is obtained by introducing a purely poloidal field \citep{Haskell:2007bh}, 
\begin{align}
B_r &= \frac{B_p \cos \theta}{\pi(\pi^2-6)} [y^3+3(y^2-2)\sin y +6y\cos y]\\
B_{\theta} &= \frac{B_p \sin \theta}{2\pi(\pi^2-6)} [-2y^3+3(y^2-2)(\sin y-y\cos y)]\\
B_{\phi} &= 0.0
\end{align}
inside the star  and
\begin{align}
B_r &= \frac{B_p R^3 \cos \theta }{r^3} \label{outB1} \\
B_{\theta} &= \frac{B_p R^3 \sin \theta}{2r^3} \label{outB2} \\
B_{\phi} &= 0 \label{outB3}
\end{align}
outside the star, where $B_p$ is the surface poloidal magnetic field strength, which we set to be $B_p = 10^{17}$ G. Such a strong magnetic field reduces the timescales allowing us to explore greater possibilities within a shorter run of the simulation. To accelerate the development of the instability, we add a small perturbation to the velocity of the fluid elements located 
at ($60^{\circ} \leq \theta \leq 120^{\circ}$) and ($7 \, \rm km \leq r \leq 9 \, \rm km $), given by:
\begin{align}
v_{\theta} &= \sqrt{\frac{15}{8\pi}}\sin\theta\sin 2\phi \\
v_{\phi} &=\sqrt{\frac{15}{8\pi}}\sin\theta\sin 2\phi\cos\theta
\end{align}
We confirm that this has no other effects apart from triggering the instability which still grows but takes longer to develop without the perturbation. Additionally, we also test another initial condition, with the same poloidal field, but a stronger toroidal component ($B_{t} = 2 \times 10^{17}$) inside the star given by

\begin{align}
B_{\phi} = B_{t} \frac{\sin y\sin \phi }{\pi},
\end{align}
and study its evolution with time.

\subsection{{Timescales}} \label{Tm}
The two important timescales in our simulation are the sound crossing time ($\tau_{c_s}$) and the {\alfven} time ($\tau_{A}$). $\tau_{A}$ is defined by
\begin{equation}
\tau_{A} = \frac{2 \rstar \sqrt{4 \pi  \langle\rho\rangle }}{\avgB}
\end{equation}
where, $\langle .. \rangle$ represent volume averaged quantities. The field evolution depends on the {\alfven} timescale while the hydrostatic equilibrium depends on the shorter sound crossing time. We chose $B_p$ such that $\tau_{c_s}\sim 0.1 \tau_{A}$.
The {\alfven} time is not a constant throughout the entire run of the simulation, instead it varies with the change in magnetic field. Initially, the magnetic field rearranges and its density becomes higher in the core of the star. Thus, $\tau_{A}$ changes in subsequent times and does not remain constant. In our simulations with $B_p = 1\times10^{17}$ G, we obtain an average $\tau_{A} = 1.3$ ms.

\subsection{{Boundary Conditions}} \label{BC}


Our objective is to understand the interior field strength and configuration of a NS if the exterior dipolar field has a given strength, inferred from observations. With this physical picture in mind, we set the exterior boundary condition of our simulation by setting the magnetic field in the $r$ and the $\theta$ directions according to equations \ref{outB1}-\ref{outB3}. At the outer ghost cells, we evaluate $B_r$ and $B_{\theta}$ at r = r[END]+$\delta$r. Here r[END] is the radial value corresponding to the end of the atmosphere, and $\delta$r is difference in radial grid spacing. We use periodic boundaries for the magnetic field along the $\phi$ direction. The velocities are all set to zero at the boundaries.

Note that to correspond to the physical prescription described above, the exterior boundary should be far from the star, where the dipolar component provides the dominant contribution to the spin-down torque. Due to numerical limitations we are far more restricted and, in practice, have to place our exterior boundary close to the star. As already mentioned we have studied two setups, with the exterior boundary at 1.2 {\rstar} and another which extends up to 2{\rstar}. We find that our results do not depend significantly on the location of the outer boundary. Extending the atmosphere farther out, in regions where higher multipoles of the field fall off much more rapidly than the dipolar component, will, thus not alter our conclusions. 

At the inner boundary in the radial direction, we set the density of the star to follow the profile given by equation \ref{rhoeq} while at the outer boundary we set the density to be $\rho_{atm}$.


\begin{figure*}
	\centering
	\begin{subfigure}{.51\textwidth}
		\centering
		\includegraphics[width=0.95\linewidth]{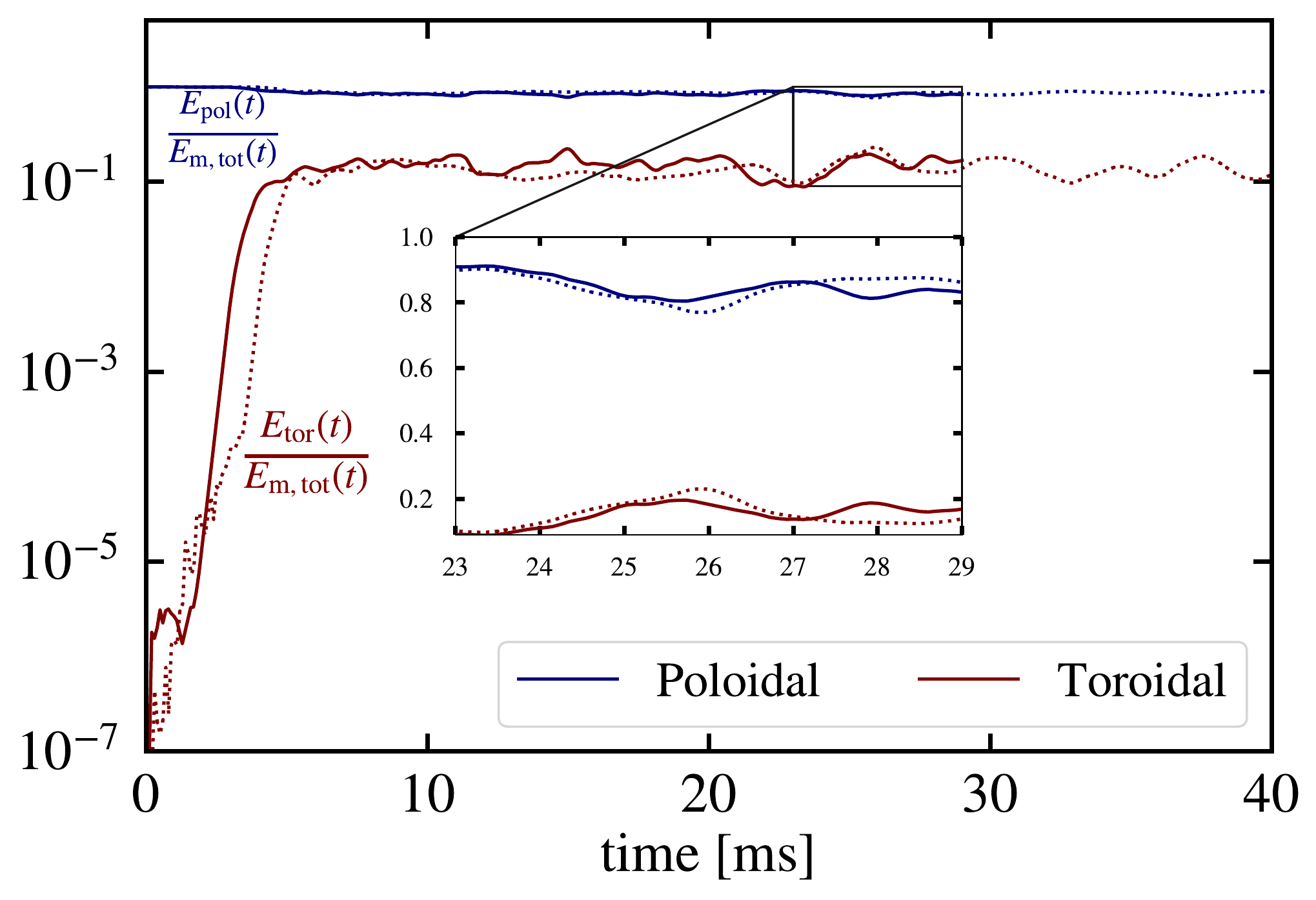}
		\caption{}
		\label{cgsbratio}
	\end{subfigure}%
	\begin{subfigure}{.51\textwidth}
		\centering
		\includegraphics[width=0.95\linewidth]{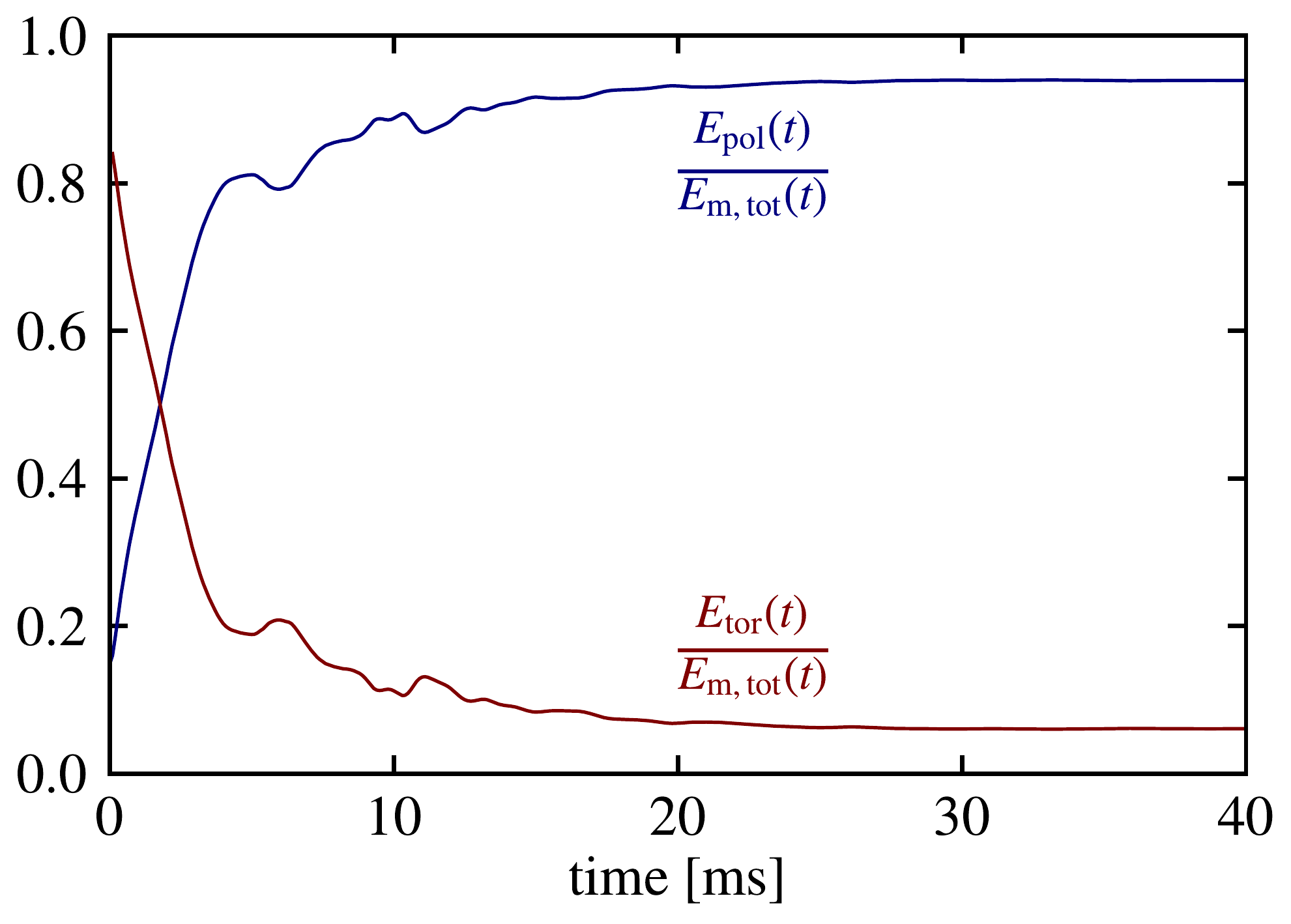}
		\caption{}
		\label{mixedfield}
	\end{subfigure}
	\caption{ (a) A comparison of the poloidal ($E_{\text{pol}}$) and toroidal ($E_{\text{tor}}$) field energies normalized by the total magnetic field energy $E_{\rm m,tot}$ at each time $t$ for our model with $\{ B_p = 10^{17}$ G, $B_t = 0$ $\}$. The atmosphere extends up to 12 km (represented by solid lines), whereas the dotted lines represent a model with an extended atmosphere till 16 km. The inset in the figure shows a linear scale comparison between the two components. (b) Evolution of the field energies starting with an initially stronger toroidal field with  $\{ B_p = 10^{17}$ G, $B_{t} = 2 \times 10^{17}$ G.$\}$ We find $E_{\rm tor} \leq 20\% \, E_{\rm m,tot}$ in both the cases.}
\end{figure*}
 
\begin{figure*}
	\includegraphics[scale=0.35]{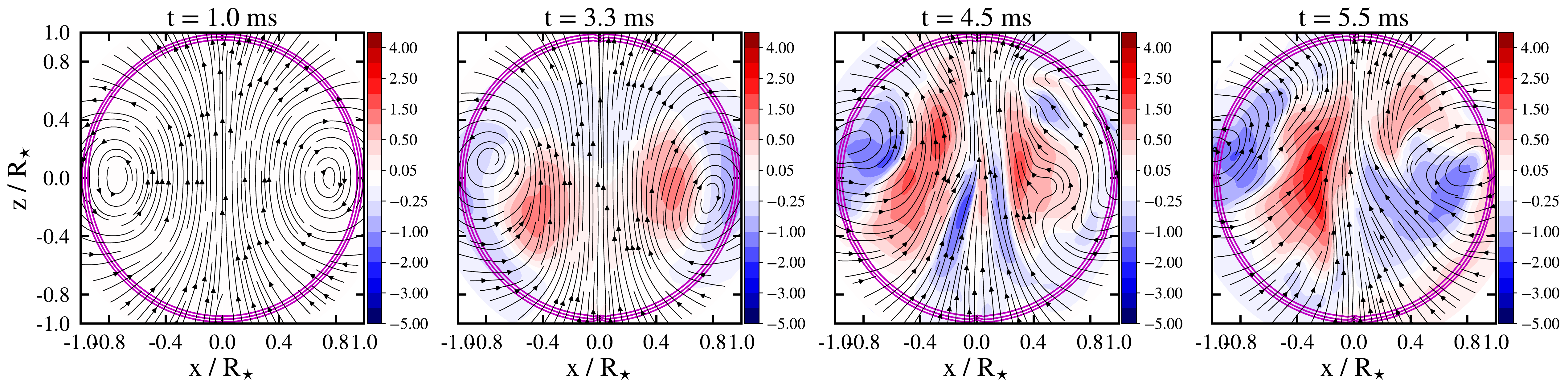}
	\label{bfieldlinesxz}
	\includegraphics[scale=0.35]{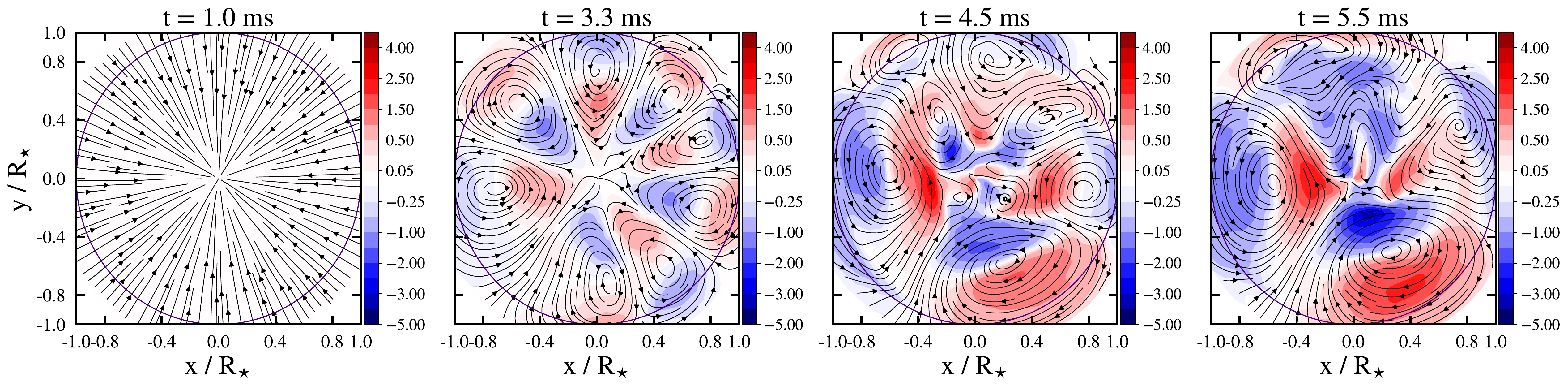}
	\caption{Snapshots of the star from meridoinal view (top) and equatorial view (bottom) showing the development of a toroidal field (color scale indicates strength of $B_{\phi}$ normalized by $2 \times 10^{16}$ G). In the top row, the purple lines show equidensity contours $\rho \in (10^{13}, 5\times10^{13}, 10^{14})$ \gpercc. The streamlines shown are the poloidal fieldlines that thread through the main body of the star. In the bottom row, the streamlines show the toroidal fieldlines and the violet contour shows the location of $\rstar$. Times of the snapshots are given as figure titles.}
	\label{bfieldlinesxy}
\end{figure*}

\vspace{0.4 cm}
\section{Results\label{results}}
\label{results}
We now move on to discuss the results of our simulations. In this section we will first present the results of our non-resistive setup, and then discuss all the results corresponding to the resistive atmosphere, in detail, in section \ref{resistivity}. Nevertheless the main conclusions are not affected by the choice of setup.
All setups, both resistive and non-resistive, are initially unstable, independently of whether we choose a purely poloidal field as initial condition or a twisted torus with a stronger toroidal component. In all cases we find that turbulence develops and the field settles to a state which is not strictly an equilibrium, but in which the non-linear saturation of the instability determines a stable average of the field strengths, such that the energy of the toroidal component is roughly $10-20\%$ of the total magnetic energy. 

\begin{figure*}
	\centering
	\begin{subfigure}{.51\textwidth}
		\centering
		\includegraphics[width=0.95\linewidth]{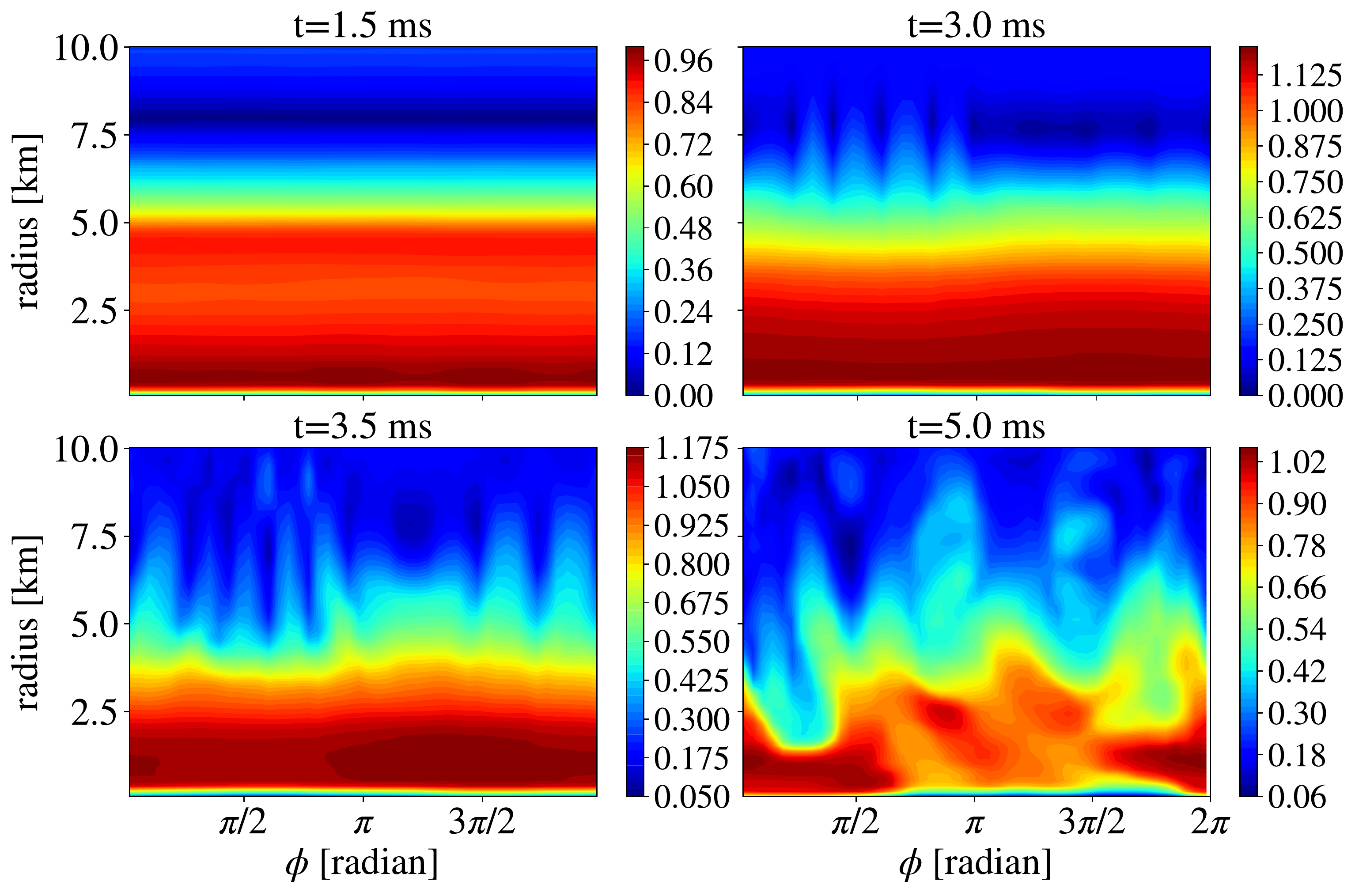}
		\caption{$\varepsilon_M$}
		\label{Babs}
	\end{subfigure}%
	\begin{subfigure}{.51\textwidth}
		\centering
		\includegraphics[width=0.95\linewidth]{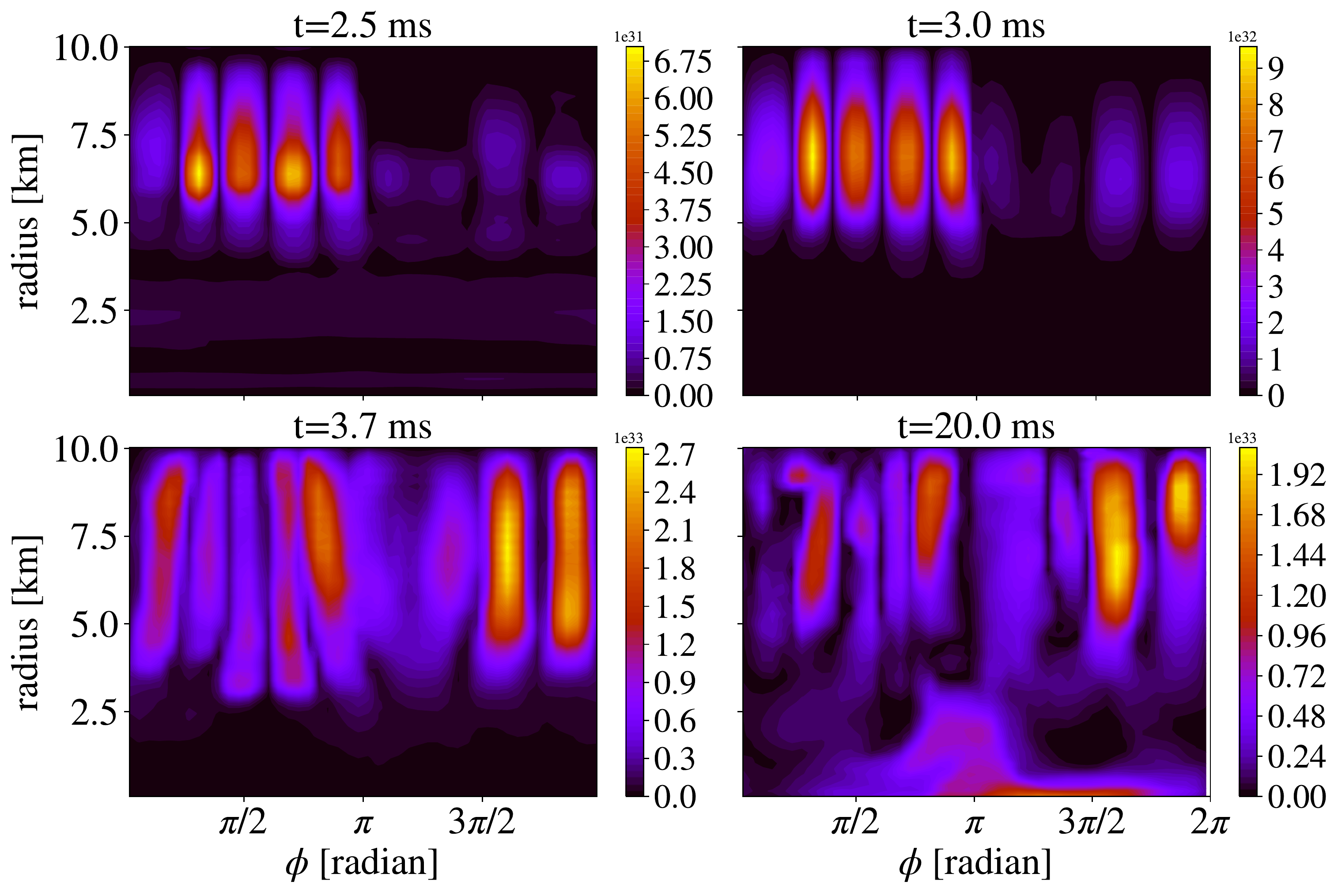}
		\caption{$\varepsilon_K$}
		\label{KEabs}
	\end{subfigure}
	\caption{Energy densities plotted in the equatorial plane of the star, i.e. at $\theta=90^{\circ}$ for different times $t$ in our simulation. At $t = 3$ ms, we see that the neutral line gets disrupted showing the formation of a strong toroidal component.}
	\label{magrphi}
\end{figure*}

\begin{figure*}
	\centering
	\begin{subfigure}{.51\textwidth}
		\centering
		\includegraphics[width=1.0\linewidth]{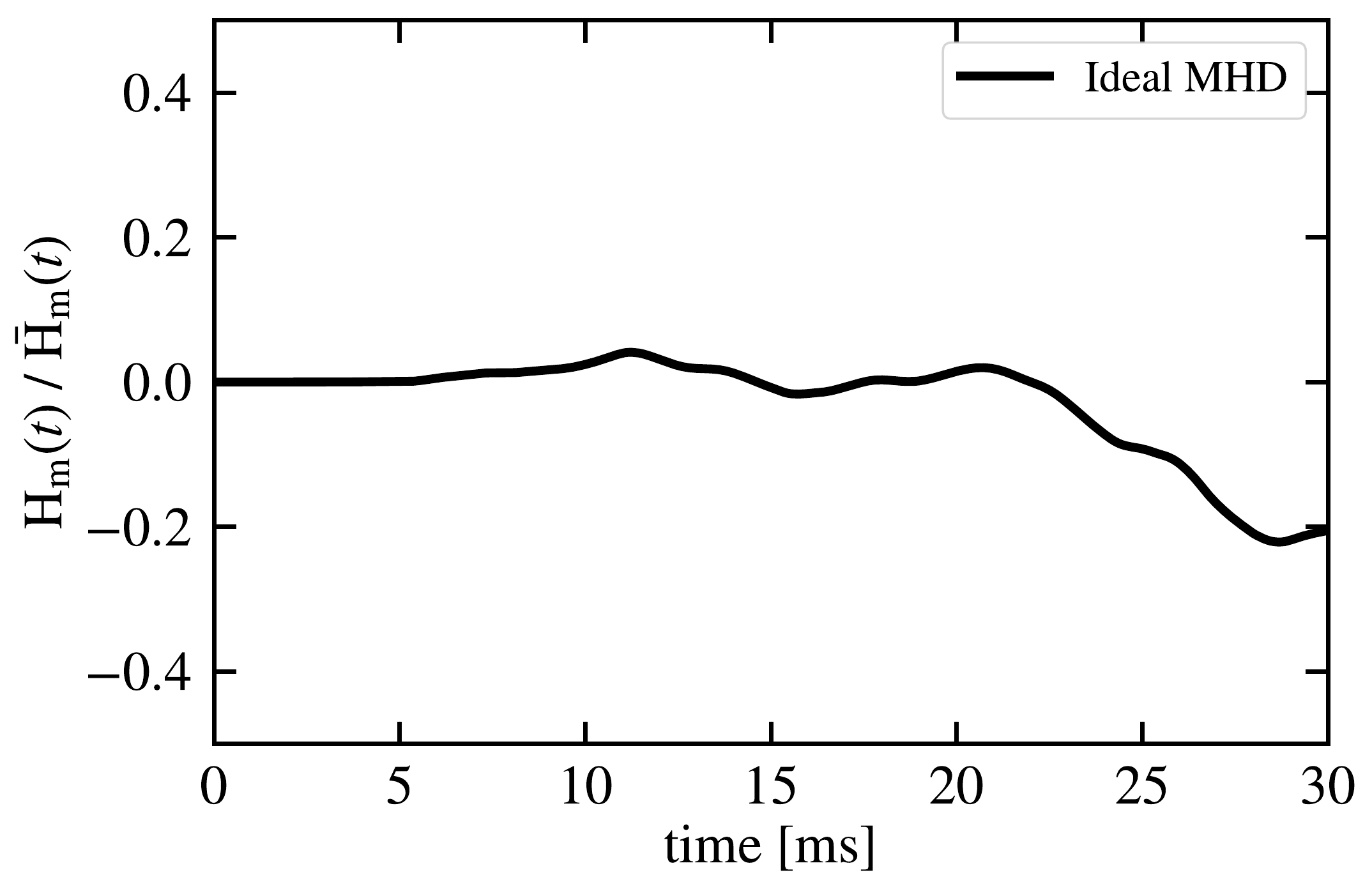}
		\caption{}
		\label{helicityplot1}
	\end{subfigure}%
	\begin{subfigure}{.51\textwidth}
		\centering
		\includegraphics[width=1.0\linewidth]{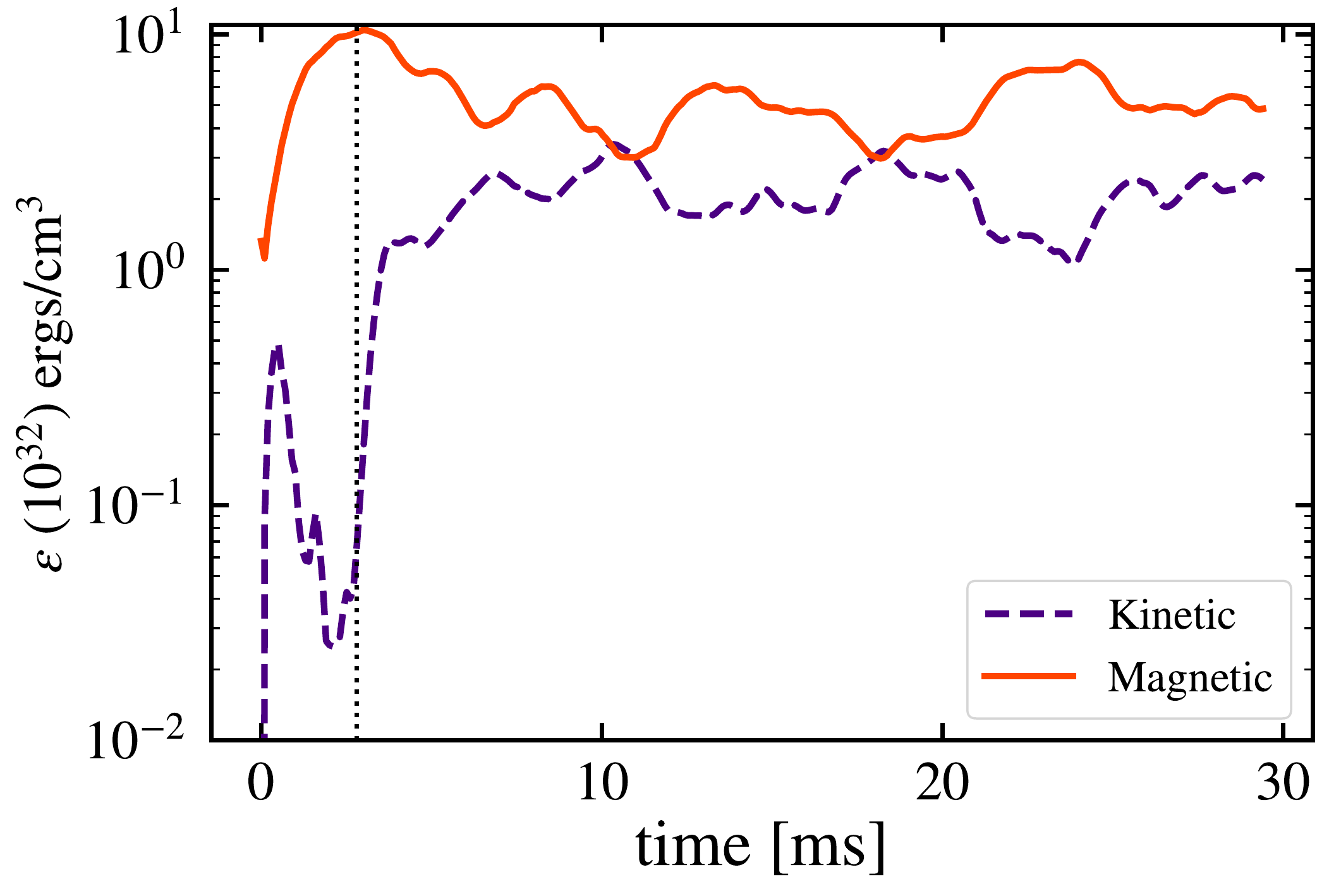}
		\caption{}
		\label{deltaBE}
	\end{subfigure}
	\caption{ (a) Time evolution of the magnetic helicity for ideal-MHD setup (i.e. resistivity $\eta_0 = 0$). (b) Average magnetic field energy density  ($\varepsilon_M$) and the average kinetic energy density ($\varepsilon_K$) normalized by $10^{32}$ ergs/cm$^3$ plotted as a function of time. After an initial transient, the onset of instability is seen at $t = 3$ ms (represented by the dotted line) when the there is a sharp rise in the kinetic energy of the system.}
\end{figure*}

\subsection{Field Configurations and Strengths}
Let us analyse in detail the evolution of the relative strengths of the poloidal and toroidal field components for our non-resistive setup.
Figure \ref{cgsbratio} shows the evolution of poloidal and the toroidal magnetic field energies normalized by the total magnetic field energy at each time for the entire run of a simulation in which the initial condition was a purely poloidal field. The toroidal component initially gains strength from the initial perturbation we gave. After 3 ms, the poloidal field becomes unstable and the toroidal component undergoes an exponential growth with its strength becoming comparable to the poloidal component. This period of exponential growth corresponds to 1 {\alfven} crossing time during which the toroidal component fully develops close to the neutral line. After $t = 5$ ms, the field reaches pseudo-equilibrium and the evolution becomes less dramatic. However, the  toroidal part remains weaker ($\leq 20 \%$ of the total magnetic energy) than its poloidal counterpart. Even with a mixed-field initial condition with a stronger toroidal component ($B_t = 2 \times 10^{17}$ G as compared to a poloidal strength of $B_p = 10^{17}$ G), we see that the final configuration settles to a weaker toroidal field as compared to the poloidal field (see figure \ref{mixedfield}). 

\begin{figure*}
	\includegraphics[scale=0.64]{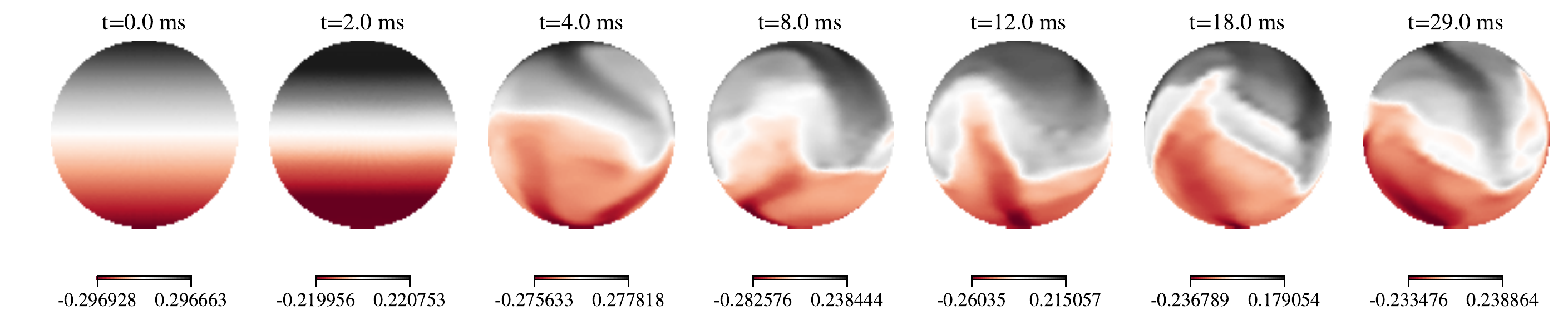}
\end{figure*}
\begin{figure*}
	\includegraphics[scale=0.64]{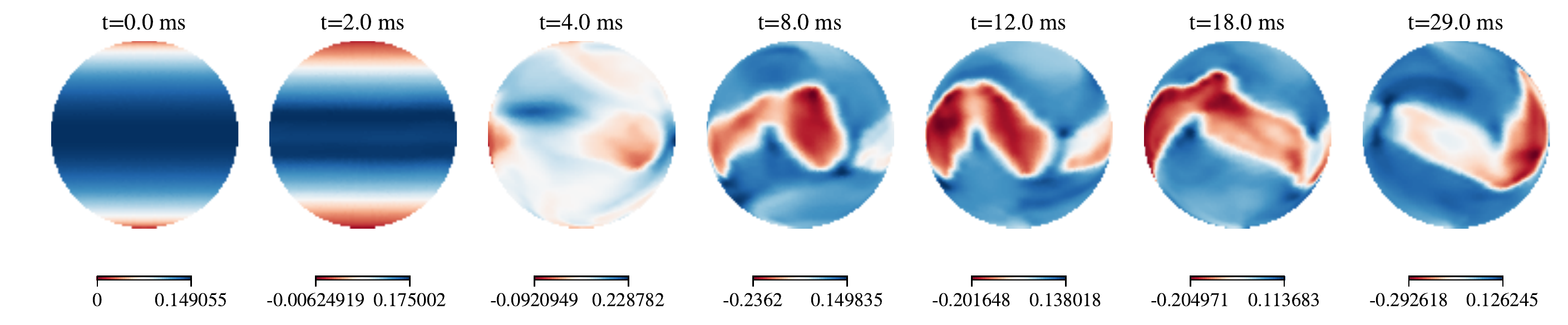}
\end{figure*}
\begin{figure*}
	\begin{tcbraster}[raster columns=4, enhanced, blankest]
		\tcbincludegraphics[scale=1.0]{FinerCGS/field0.pdf}
		\tcbincludegraphics[scale=1.0]{FinerCGS/field3.pdf}
		\tcbincludegraphics[scale=1.0]{FinerCGS/field10.pdf}
		\tcbincludegraphics[scale=1.0]{FinerCGS/field20.pdf}
	\end{tcbraster}
	\caption{ Surface distribution of $B_r$ (top) and $B_{\theta}$ (middle) at various times t, given as figure titles. The colorscale (normalized by $10^{17}$ G) shows the strength of the field. (bottom) 3-D visualization of the magnetic field configuration at times t=0 ms, t=3 ms, t=10 ms and t=20 ms (from left to right).}
	\label{3dfield}
\end{figure*}

Figure \ref{bfieldlinesxy} shows snapshots of the formation of the toroidal component in the meridional (x-z) plane and the equatorial (x-y) plane, for an initially poloidal field. The colors show the strength of $B_{\phi}$ only.
The region inside the star, where the fieldlines close, moves in and out during the initial stage when the fluid starts readjusting to the changing magnetic field. The null line seems to move outwards during the evolution. As pointed out by \cite{GlampLasky2015}, closed fieldlines in the core of the star cause it to be magnetically decoupled with the rest of the star by developing a velocity lag between these regions. During the initial stage, the field can be treated as a linear perturbation on top of a stationary background field. The nonlinear terms starts to dominate after the onset of the ``kink''-instability at $t \sim 3$ ms. The dynamics thereafter change rapidly, breaking the axisymmetry and the field inside the star attends a complex geometry with the mixed-field configuration. However, there is a caveat. The nonlinear terms may have an initially stronger role because of our strong field. The instability is, however, inherently nonlinear, and will in general grow until these nonlinear terms block it. The toroidal component creates vortex-like structures (shown in the lower-panel of figure \ref{bfieldlinesxy}) in order to conserve the magnetic helicity \citep{Ciolfi:2011xa} which is initially zero due to our choice of a purely poloidal field as initial condition.
These structures show higher-order modes (see figure  \ref{bfieldlinesxy} at $t=3.3$ ms) which are replaced by lower-order modes at later stages during the evolution. The presence of the kink-instability is visible in our simulation in figure \ref{Babs}, where the absolute value of the magnetic field strength is plotted on the equatorial plane of the star. The deep blue line feature at $r \sim 8$ km at $t=1.5$ ms shows the location of the neutral line. This gets distorted and small lumps are seen at $t = 3$ ms which evaporate thereafter. Similarly, we plot the kinetic energy of the star on the equatorial plane (figure \ref{KEabs}) and note the different modes of oscillations present in our simulation. Higher order modes are visible at $t = 2.5$ ms, and start coupling with each other at later stages as seen at $t = 20$ ms.

The magnetic helicity $\rm H_m$ measures the amount of `twist' in the magnetic field and is given by:
\begin{eqnarray}
{\rm H_{\rm m}} = \int_V \bm{A} \cdot \bm{B} \, dV
\label{hell}
\end{eqnarray}
where $\bm{A}$ is the magnetic vector potential. The helicity in equation (\ref{hell}) is a conserved quantity in ideal MHD, with a non-zero value generally linked to non-ideal effects such as reconnections. Figure \ref{helicityplot1} shows the time evolution of the quantity $\rm H_m/\rm \bar{H}_{\rm m}$, where the magnetic helicity is normalized by $\rm \bar{H}_{\rm m}$ = $0.5\times $ E$_{\rm m,tot} \times 0.8{\rstar}$, where $\rm E_{\rm m,tot}$ is the volume integrated total magnetic energy. We remark that the choice of gauge for $\bm{A}$ is irrelevant in this case. Initially, the helicity remains zero until there is axisymmetry in our simulation. However, as the star tries to reach an equilibrium, the helicity becomes non-zero.

We will see in detail section \ref{turbulence} that this is linked to the development of turbulence, in which following the initial development of kinetic helicity after the instability, an inverse cascade takes helicity from the resistive small scales to larger scales. This is expected as the system attempts to conserve helicity by transferring it from the small scale turbulent field to the larger scale field, thus moving it further from the resistive scale (\cite{biskamp2003}).

\begin{figure*}
	\centering
	\begin{subfigure}[b]{0.5\textwidth}
		\centering
		\includegraphics[width=1.0\linewidth]{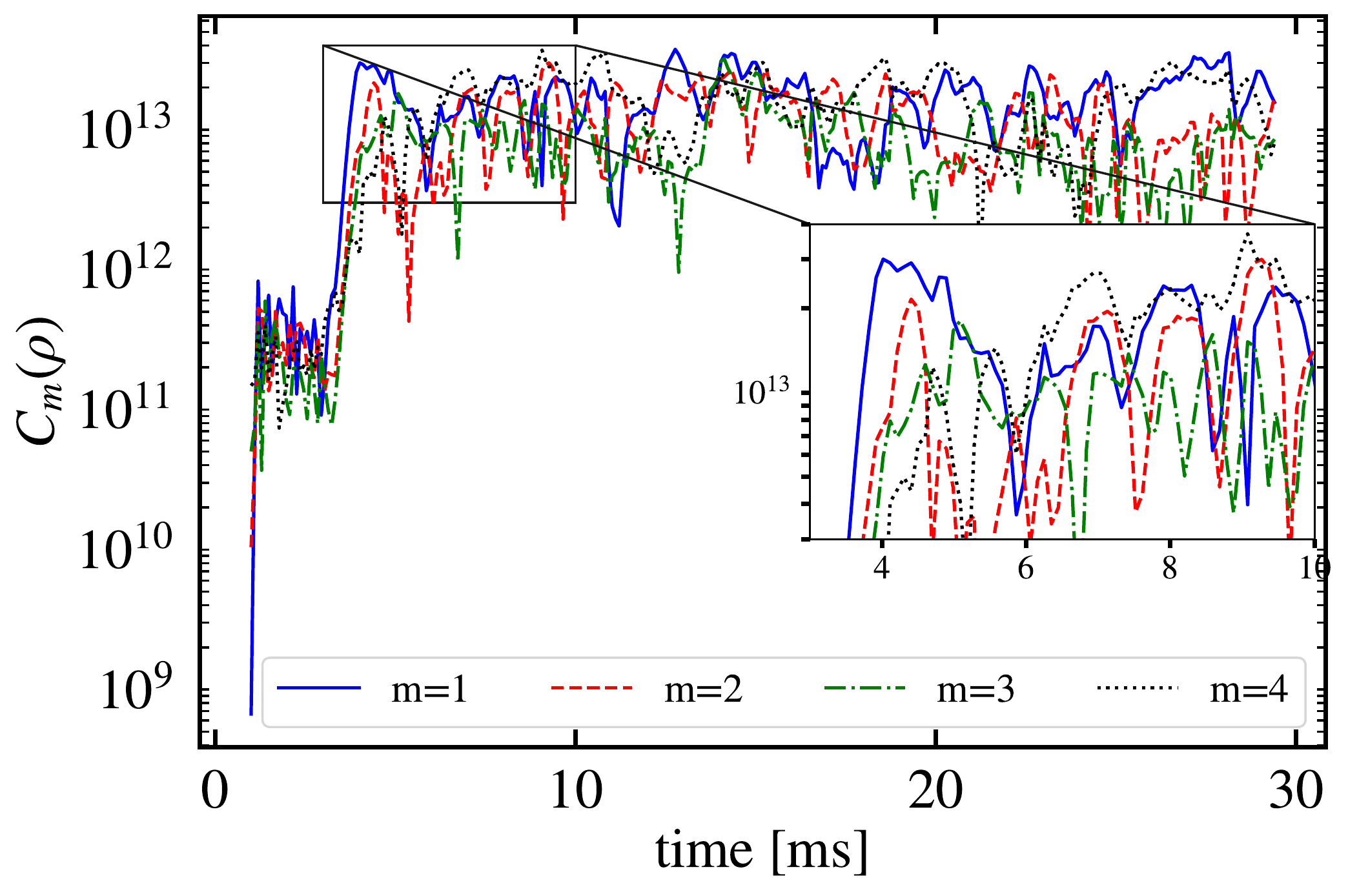}
		\label{moderho} 
	\end{subfigure}%
	\begin{subfigure}[b]{0.5\textwidth}
		\centering
		\includegraphics[width=1.0\linewidth]{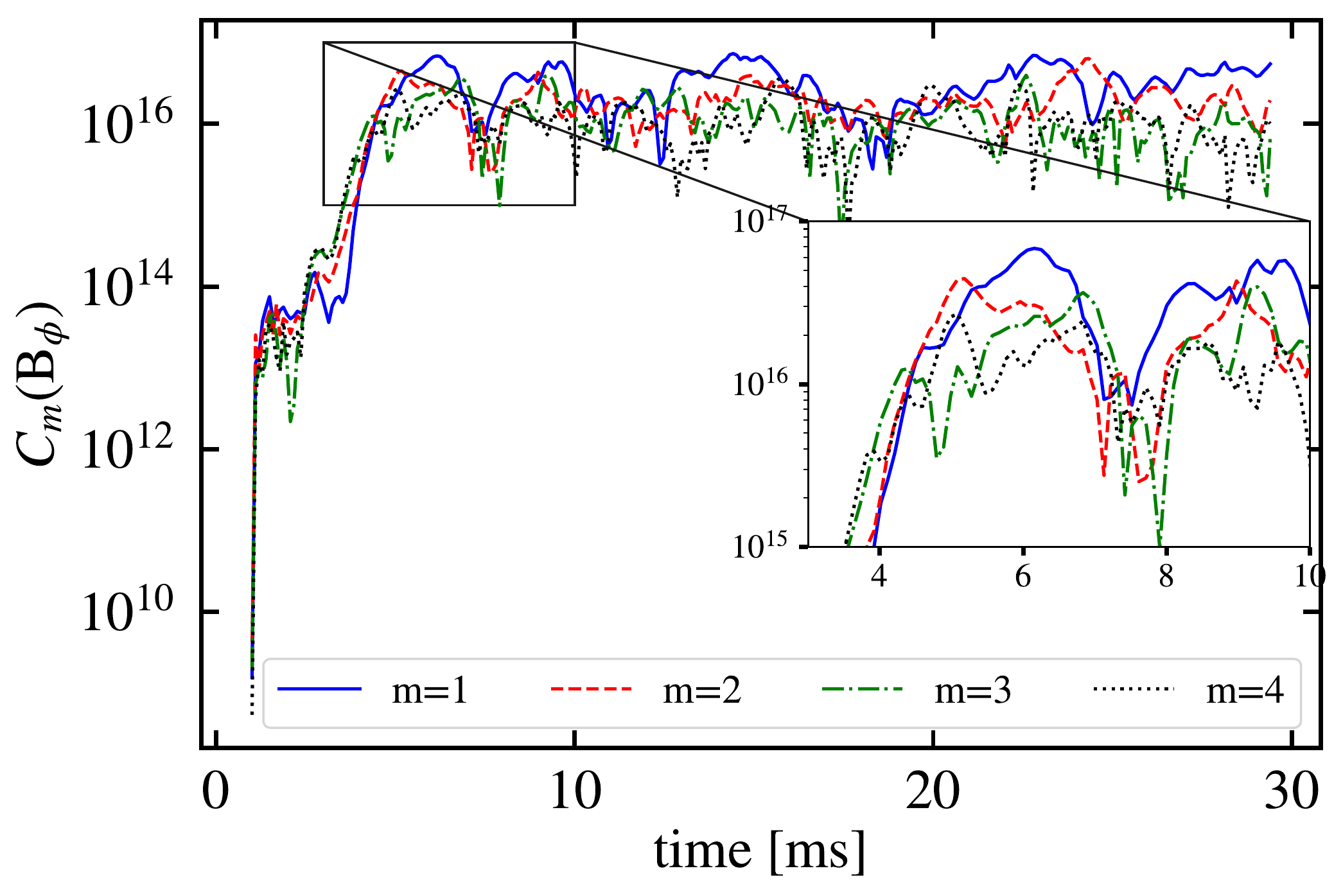}
		\label{modebphi}
	\end{subfigure} 
	\caption{Fourier decomposition of $\rho$ (left) and $B_{\phi}$ (right) into different azimuthal modes $m \in (1,2,3,4)$ as a function of time for $B_p = 10^{17}$ G. The inset shows that the lower order modes for $\rho$ grow faster compared to the higher order modes however the reverse is observed for $B_{\phi}$ where the higher order modes grow faster. An exponential growth is seen in both the quantities which saturates after few {$\alfven$} crossing times when the star attends a pseudo-equilibrium state.}
	\label{modes1}
\end{figure*}

\begin{figure*}
	\centering
	\begin{subfigure}[b]{0.48\textwidth}
		\centering
		\includegraphics[width=1.0\linewidth]{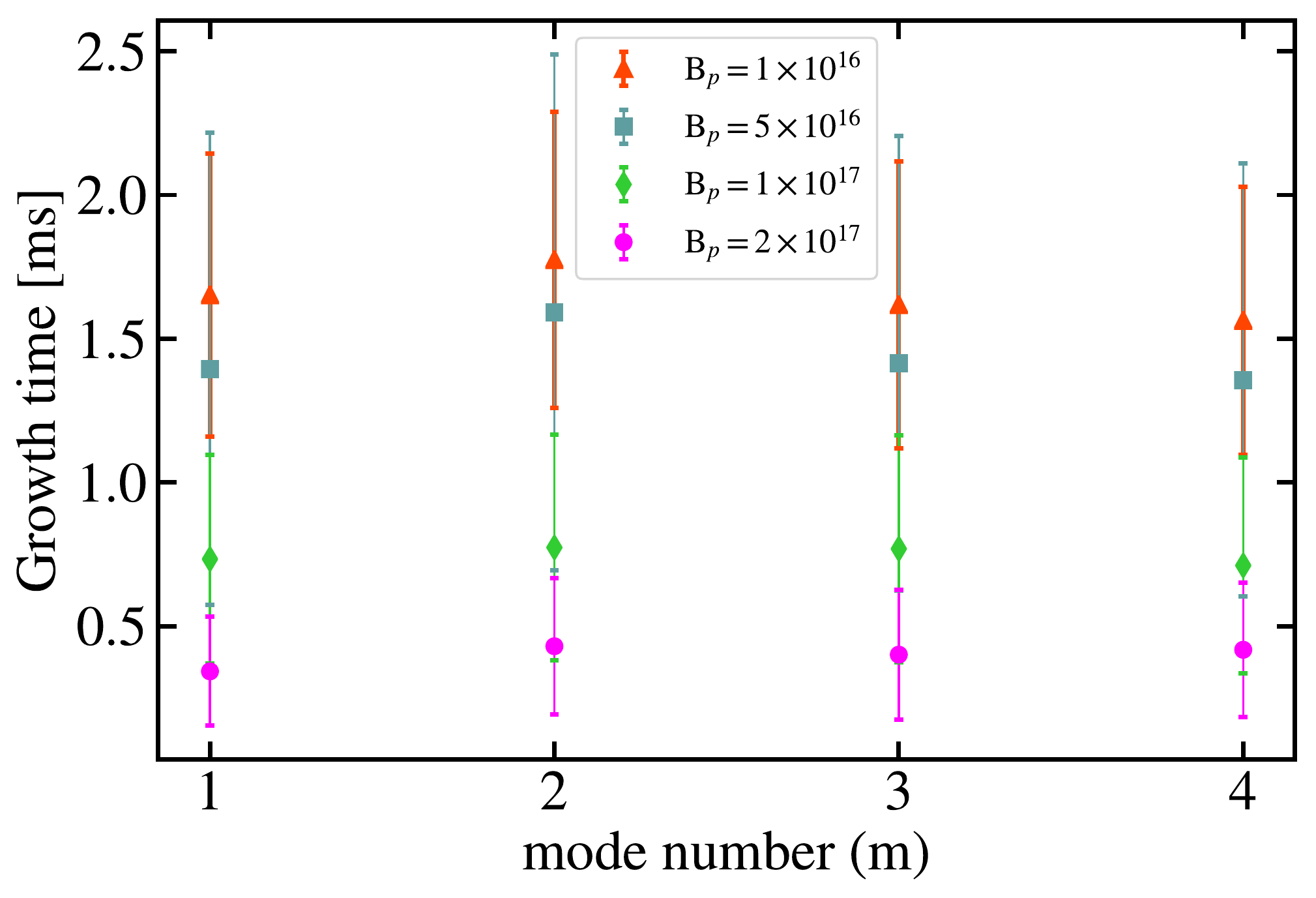}
		\label{fig:cm3}
	\end{subfigure}
	\begin{subfigure}[b]{0.48\textwidth}
		\centering
		\includegraphics[width=1.04\linewidth]{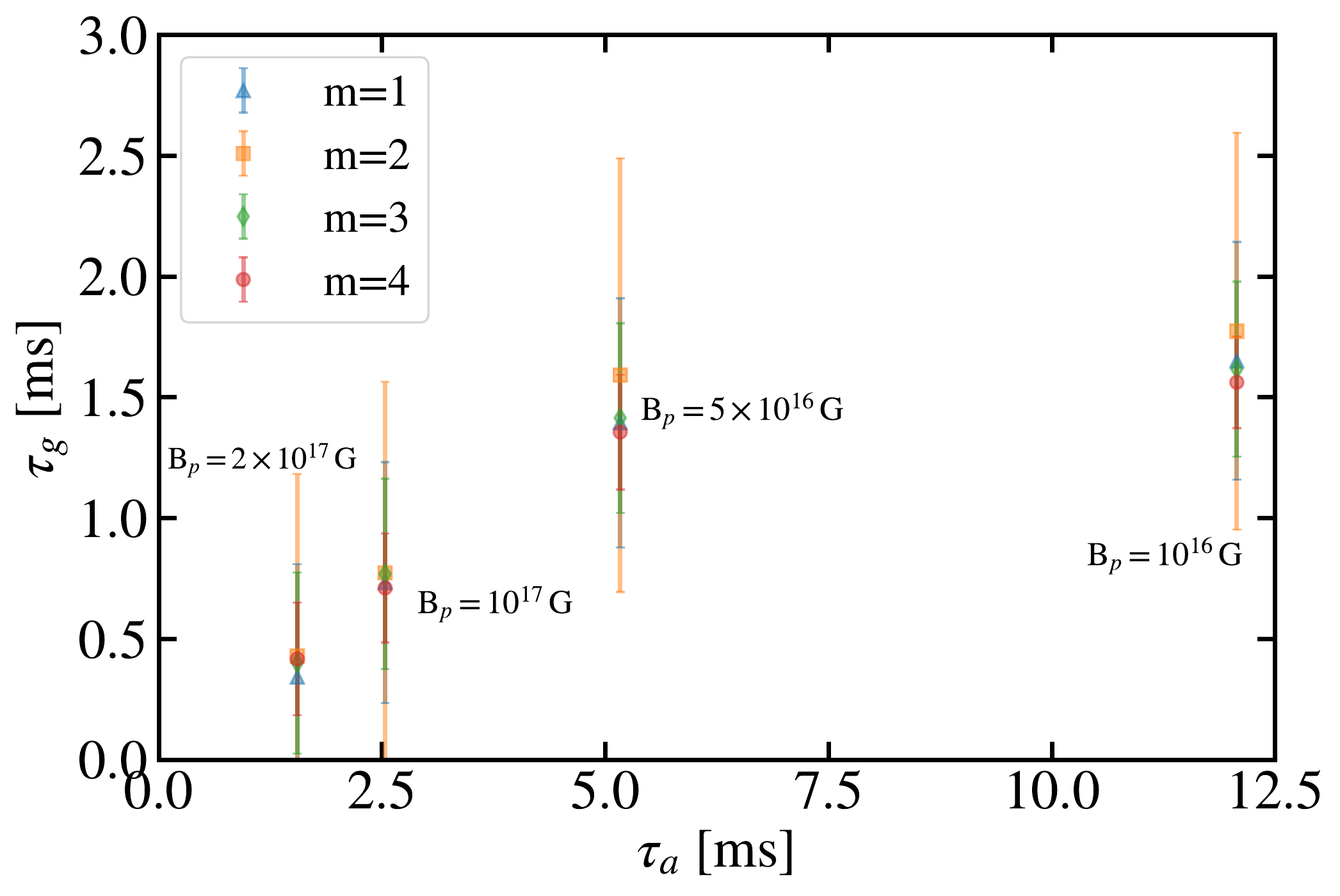}
		\label{fig:cm4}
	\end{subfigure}
	\caption{(left) shows the growth times for different modes with varying surface magnetic field strengths. (right) shows correlation between the growth times and the {$\alfven$} crossing time. Our results are consistent with the expected linear relation between two quantities, but given that our error bars are large we do not present a fit to the data.}
	\label{modes2}
\end{figure*}

The atmosphere also plays a role in governing the internal dynamics of the field. We explore its effect by running simulations with different values of $\rho_{\text{atm}}$ and find that an atmosphere with higher-density fluid (which allows the star to lose more magnetic energy) causes a relatively weaker toroidal field as compared to an atmosphere with lower-density fluid. Extending the atmosphere up to a larger distance also does not influence much the overall growth of field-energies. This is shown as dotted lines in figure \ref{cgsbratio}.
As the influence of higher order multipoles is stronger close to the surface, the fact that the results are mostly unaffected by the position of the boundary within a few stellar radii gives us a degree of confidence that these are a good approximation to the physically realistic situation, in which the dipolar field component is inferred further out at the light cylinder.

Figure \ref{deltaBE} shows the variation of the volume averaged magnetic energy density ($\varepsilon_M = (B_r^2+B_{\theta}^2+B_{\phi}^2)/8\pi$) and the volume averaged kinetic energy density ($\varepsilon_K = \rho(v_r^2+v_{\theta}^2+v_{\phi}^2)/2$) with time. A peculiar feature of our evolutions is the initial rearrangement of the poloidal field, due to our choice of initial condition in which the field is stronger in the outer core. This leads to a rapid initial readjustment which takes it to a more stable configuration in which stronger near the center of the star and weaker in the exterior regions. As a consequence at $t \sim 1$ ms, there is a small peak in $\varepsilon_K$ which the fluid gains in response to the initial readjustment of the magnetic field. Furthermore when we take a volume average of the field, this rearrangement is visible as a sudden rise in $\varepsilon_M$, as the field becomes stronger in the interior region over which we integrate, \textcolor{red}{as seen in} the first few milliseconds (figure \ref{deltaBE}). Following this initial transient the field settles down, until it is  affected by the onset of the ``kink'' instability at $t=3$ ms (after $\sim 1 \alfven$ crossing time) when the magnetic energy falls and the kinetic energy rises sharply. This is ascribed to the conversion of magnetic energy to kinetic energy (see also \cite{Lasky:2012ju}). Finally we note that the presence of the varicose mode, in which the flux tube near the neutral line undergoes a change in cross-sectional area, is difficult to observe in 3D visualization of our simulation (see figure \ref{3dfield}) because our ${\alfven}$  crossing time is small and the instability growth time is thus too rapid. However the kink-instability is somewhat visible at t=3 ms. 
\begin{figure*}
	\centering
	\begin{subfigure}{.51\textwidth}
		\centering
		\includegraphics[width=1.0\linewidth]{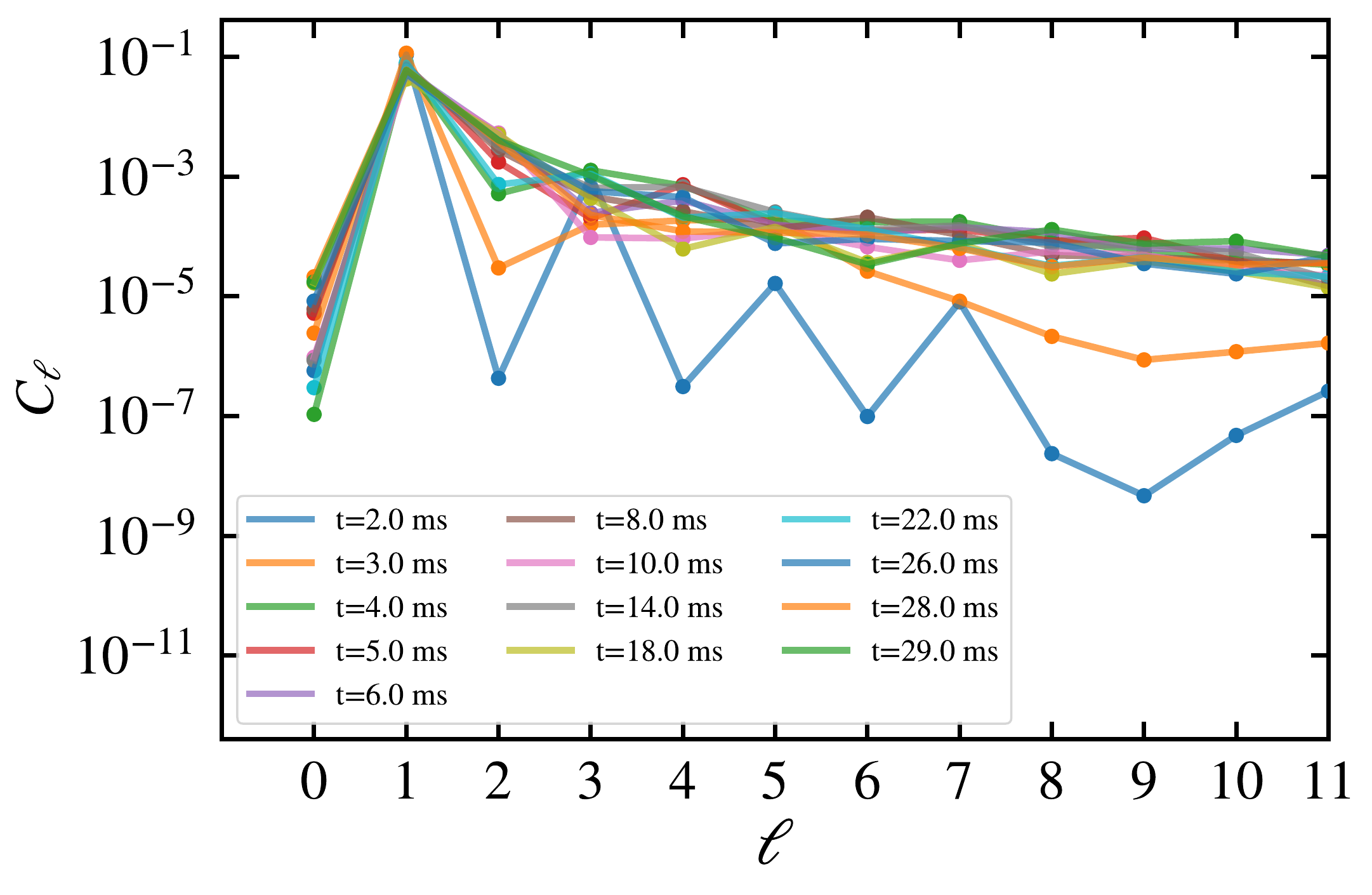}
		\caption{}
		\label{powerbr}
	\end{subfigure}%
	\begin{subfigure}{.51\textwidth}
		\centering
		\includegraphics[width=0.97\linewidth]{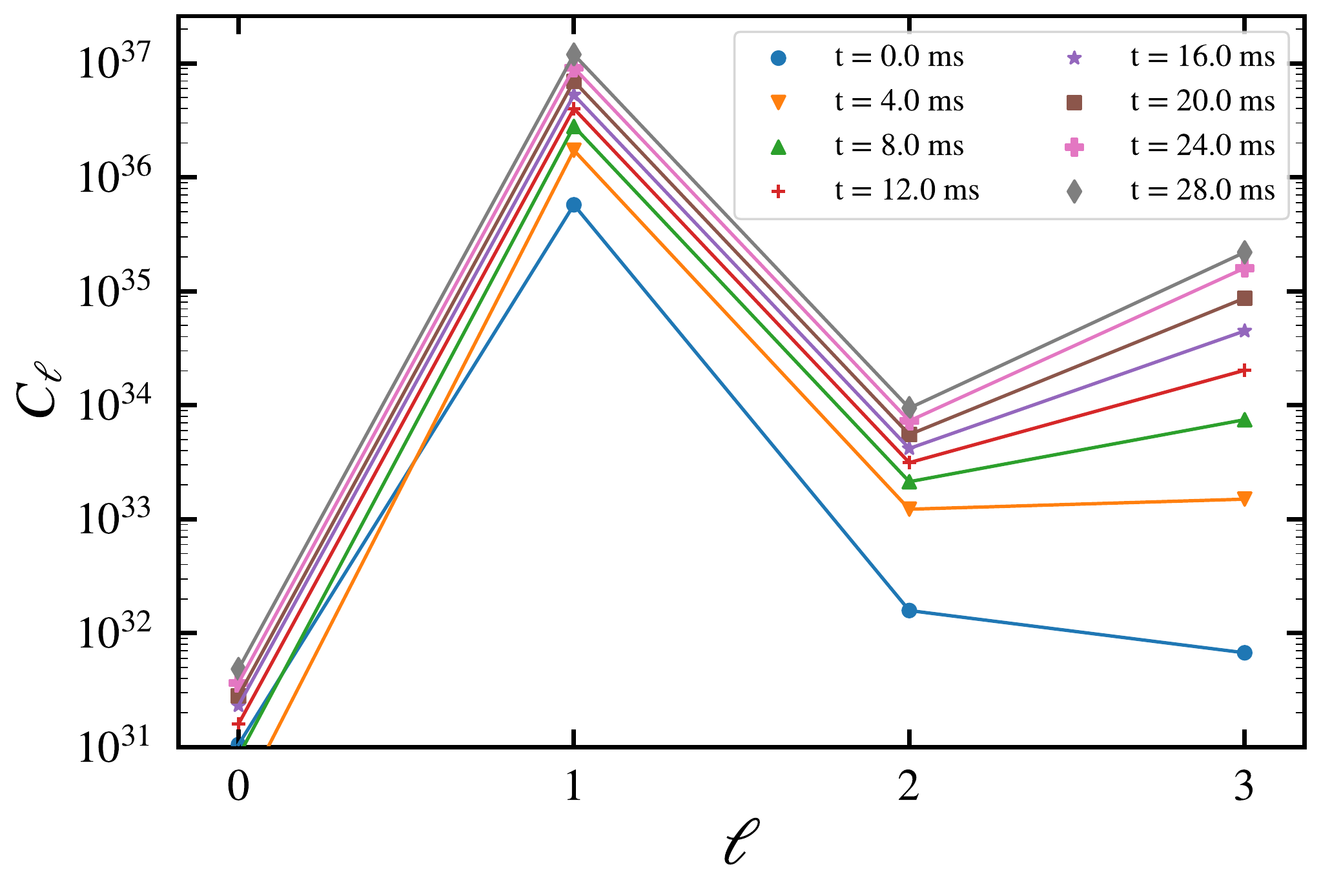}
		\caption{}
		\label{powervect}
	\end{subfigure}
	\caption{ (a) Angular power spectrum for the scalar field $B_r$ calculated at $r= {\rstar}$ plotted as a function of multipoles. (b) Power spectrum calculated for the magnetic field by decomposing it into vector spherical harmonics. The colors show different times in our simulation.}
	\label{powerspec}
\end{figure*}

\subsection{Growth Times}
We Fourier decompose $B_{\phi}$ and $\rho$ into different modes $m$ and calculate the complex weighted averages as prescribed in \cite{zink2007,Lasky:2011un} given by

\begin{eqnarray}
C_m(f) = \int_0^{2\pi} f(\bar{\omega},\phi,z=0)e^{im\phi}d\phi
\label{cm}
\end{eqnarray}
where $\bar{\omega}=\sqrt{x^2+y^2}=0.8\rstar$ lies in the equatorial plane of the star, and $f \in (\rho, B_{\phi})$. Since equation \ref{cm} results in a complex number, we take the modulus to obtain $C_m({f})$. Figure \ref{modes1} shows the modal structure of the instability. All the different modes $m \in (1,2,3,4)$ considered show the presence of the instability as each one grows exponentially by three-four orders of magnitude in one {\alfven} crossing time before settling down to a pseudo-equilibrium state. For $C_m(\rho)$, visually, the lower-order modes grow faster, i.e. $m = 1$ mode grows faster than $m = 2$, and so on, whereas for $B_{\phi}$, we see that the higher-order modes grow faster, i.e. $m = 3$ mode grows faster than $m = 2$, and so on.

Following \cite{Lasky:2011un}, the instability growth time for a particular $m$ during the exponential phase is defined by

\begin{eqnarray}
\tau_{g} = \frac{\Delta t}{\Delta \text{ln}[C_m(B_{\phi})]}
\label{tau_g}
\end{eqnarray}

Figure \ref{modes2} shows the growth times for the different modes with varying surface magnetic field strength. In calculating equation \ref{tau_g}, we do not adopt a single point, rather we take different realizations during the exponential phase and then calculate the mean and standard deviation of growth times. We find that the $\tau_{g}$ for the various modes are not significantly different from each other unlike the prediction by \cite{Tayler57} where higher-order modes have a shorter growth time. It should be noted that our field strength $B_p$ corresponds to a weaker ${\avgB}$ which is otherwise used in the literature to calculate $\tau_{A}$. From figure \ref{modes2}, we find that the growth time scales approximately with the chosen range of magnetic field, although the large error bars do not allow to accurately test the scaling.

\subsection{Power Spectrum}

We use the {\tt{healpy}} modules to calculate the angular power spectrum of $B_r$. Any scalar function defined on a sphere can be expanded into spherical harmonics. Given a map, the angular power spectrum is calculated using

\begin{eqnarray}
C_{\ell} = \frac{1}{2\ell+1} \sum_{m=-\ell}^{\ell}\langle|a_{\ell m}|^2\rangle
\label{power}
\end{eqnarray}
where,
\begin{eqnarray}
a_{\ell m} = \int d\bm{u} B(\bm{u})Y^{\star}_{lm}(\bm{u})
\end{eqnarray}

$B(\bm{u})$ is a scalar field depending on the angular coordinates $\bm{u}$ and $Y^{\star}_{lm}$ is the complex conjugate of the spherical harmonics. Figure \ref{powerbr} shows that $\ell=1$ contributes maximum to the total power in $B_r$, followed by $\ell = 2$ and $\ell=4$ respectively. In figure \ref{powervect}, we decomposed the magnetic field into vector spherical harmonics and calculated the power according to equation \ref{power} for multipoles $\ell \in (0,1,2,3)$. In this case, the $a_{\ell m}$'s were calculated using the definition of vector multipole moments (see for e.g. \cite{Barrera_1985}). Initially, the field is dipolar. However, the higher-order multipoles gain power with time and the field structure becomes complex. This can be approximately seen in figure \ref{3dfield} where the field configuration evolves, and the neutral line migrates. We note, however, that the timescale on which this tilt occurs is dictated by numerical dissipation, and faster than would be expected in a realistic NS.

\vspace{0.4 cm}

\section{Effect of Resistivity \label{resistivity}}
\label{resistivity}
\begin{figure*}
	\centering
	\begin{subfigure}{.52\textwidth}
		\centering
		\includegraphics[width=1.0\linewidth]{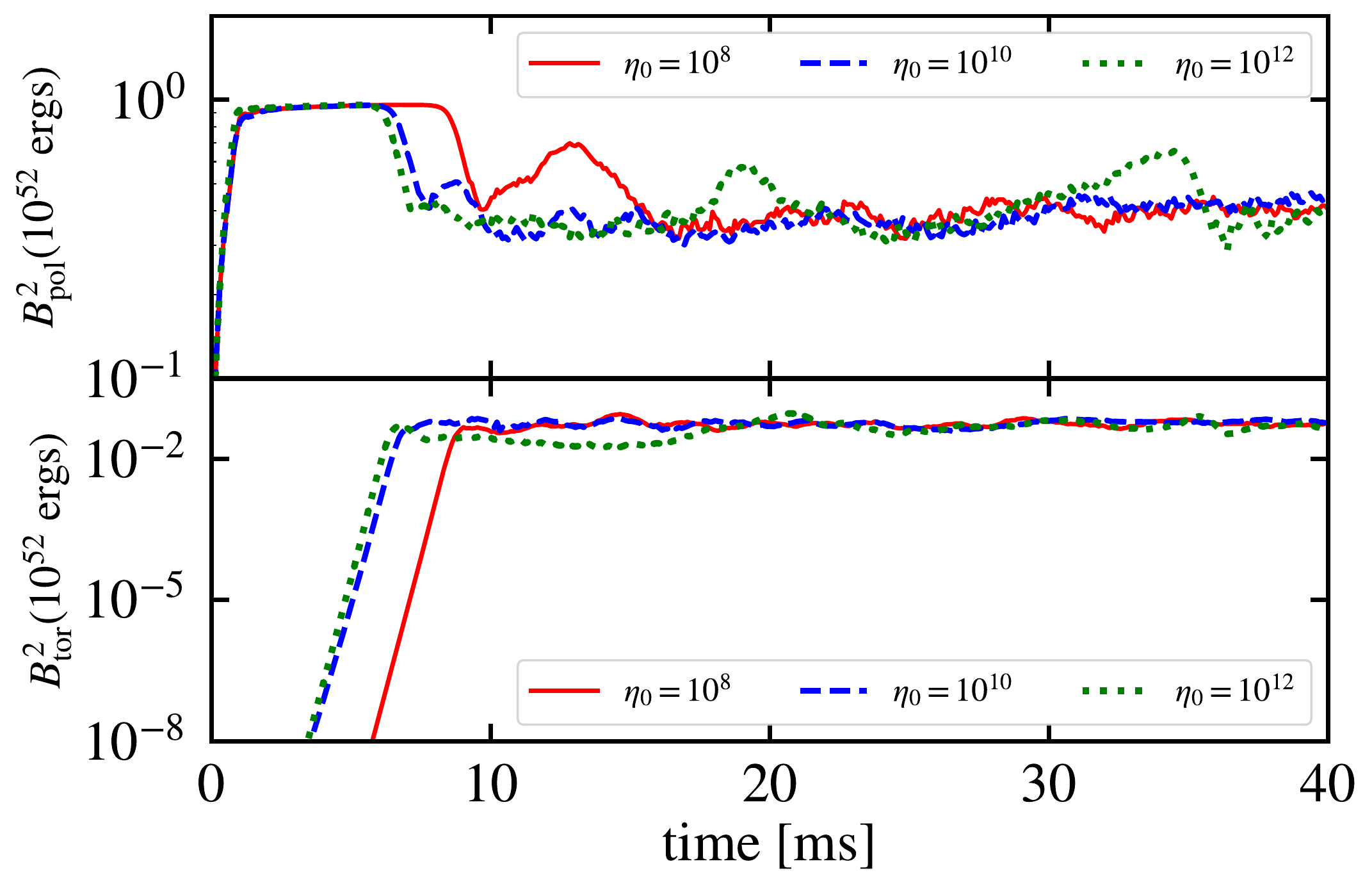}
		\caption{}
		\label{resbcomparison}
	\end{subfigure}%
	\begin{subfigure}{.52\textwidth}
		\centering
		\includegraphics[width=0.96\linewidth]{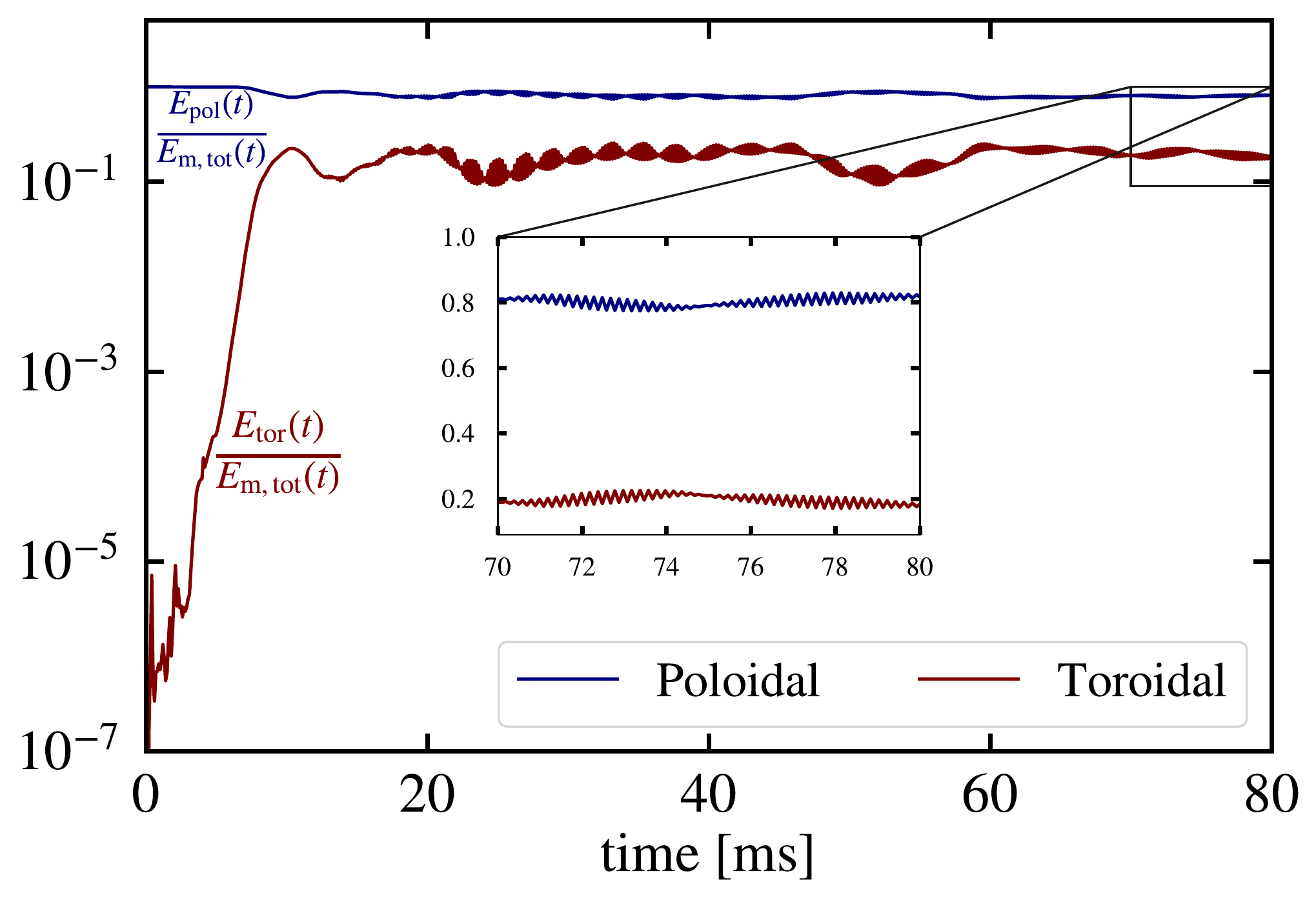}
		\caption{}
		\label{resbenergy}
	\end{subfigure}
	\caption{(a) Poloidal and Toroidal magnetic field energies as a function of time for three different values of $\eta_0$. (b) Long-term evolution for the poloidal and toroidal field energies, normalized by the total magnetic energy, for the resistive atmosphere setup. The inline plot shows that $E_{\rm tor} \sim 20 \% \, E_{\rm m,tot}$.}
\end{figure*}

\begin{figure*}
	\centering
	\begin{subfigure}{.51\textwidth}
		\centering
		\includegraphics[scale=0.42]{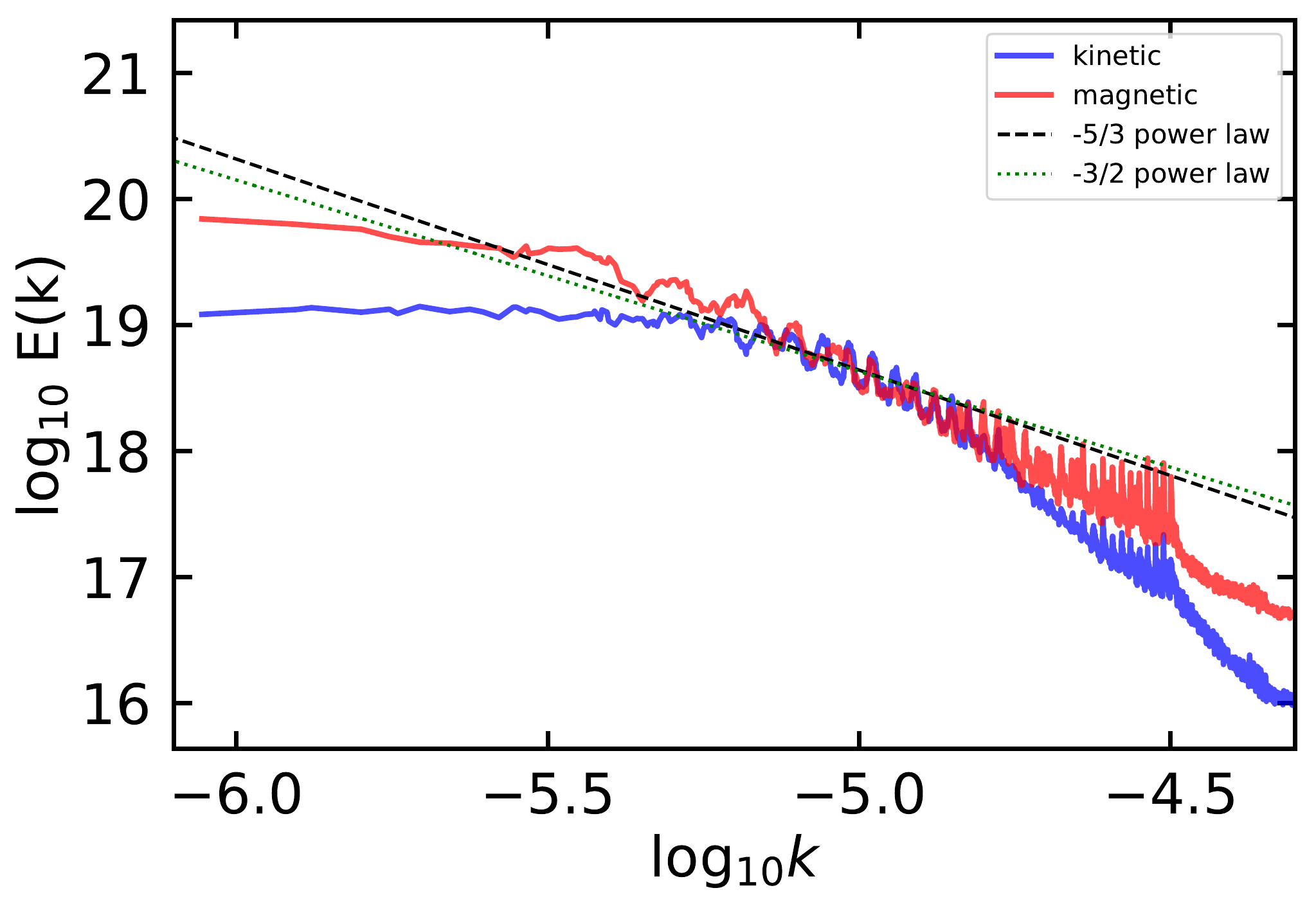}
		\caption{}
		\label{Kol}
	\end{subfigure}%
	\begin{subfigure}{.51\textwidth}
		\centering
		\includegraphics[scale=0.42]{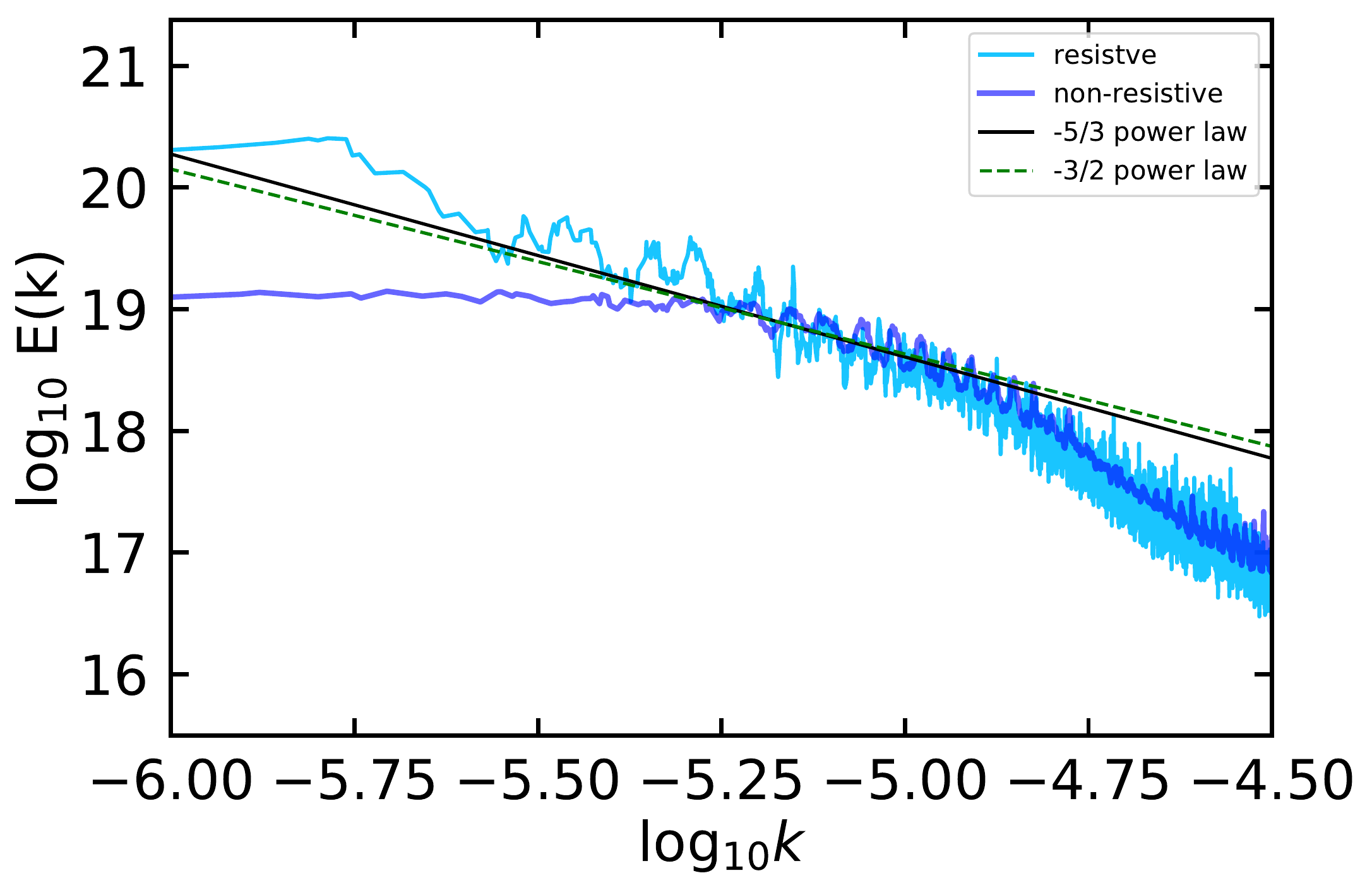}
		\caption{}
		\label{Kolres}
	\end{subfigure}
	\caption{(a) Kinetic and magnetic energy spectra for our non-resistive setup. (b) A comparison of the kinetic energy spectrum between our resistive and non-resistive setups. In both the plots, the black dashed line shows -5/3 power law (Kolmogorov) while the green dotted line shows -3/2 power law (Iroshnikov-Kraichnan). We do not have a sufficiently high resolution to distinguish between the two spectra, but our results are consistent with a Kolmogorov spectrum, as found by previous MHD simulations, as described in the text.}
\end{figure*}

In the previous section we have considered a non-resistive setup in ideal-MHD, which reflects the expectation that the NS interior is a highly conductive medium (at least for young NSs and on the dynamical timescales we are interested in, over which mechanisms such as the Hall effect or Ohmic dissipation do not have time to act). In practice, however, this approximation breaks down close to the surface of the NS as the density decreases, and resistive effects play an important role in
the long term evolution of the magnetic field (for a recent review see \citealt{PonsReview}). As one moves further towards the exterior, a low density plasma is thought to surround the NS, in which now
the tenuous fluid is tied to the magnetic field and the force-free approximation is generally used to understand the dynamics of the magnetosphere and the emission properties of the star \citep{GJ69,Spitkovsky06, Philippov18}. 

In practice, most numerical studies of fields in NS interiors have approximated the exterior plasma in terms of an atmosphere with a resistivity, mostly to prevent shocks at the stellar boundary which otherwise lead to numerical instabilities. Although our non-resistive setup is stable for a dense enough atmosphere, we nonetheless explore the effects of resistivity on the simulation, in order to investigate if any substantial differences arise. We use a profile given by $\eta(r) = 0 \, \text{if } r < \, 0.9\rstar $,
otherwise, $\eta(r) = \eta_0  \, \text{if }  r \geq \, 0.9\rstar $, where $\eta_0$ is a constant. 

The diffusion timescale, defined as $\tau_d = {\rstar^2}/\eta$, is larger than the {\alfven} crossing time ($\tau_d \geq 10 \,\tau_{A})$. We choose a value of $\eta_0$ such that this condition is satisfied. This leads us to the following relation $\eta_0 \leq {\rstar}{\avgB}/\sqrt{4\pi {\avgrho}} = 10^{12}$ {\resunit}. We set $\eta_0 = 10^8$ {\resunit} in our simulation. The above choice of the profile maintains the ideal-MHD condition in the bulk of the star. We explore different values of $\eta_0 \in \{10^{8}, 10^{10},10^{12}\}$. Additionally, we extend our atmosphere up to 20 km ($= 2 {\rstar}$). Figure \ref{resbcomparison} shows the long-term evolution of the poloidal and toroidal field energies (both normalized by the total magnetic field energy at each time) for a model with $\eta_0 = 10^{8}$ {\resunit}. Here again, we find that $E_{B_{\phi}} \leq 20 \%$ of E$_{\rm m,tot}$. The value of $\eta_0$ mostly modifies the timescales in our simulation as the onset of instability changes as illustrated in figure \ref{resbcomparison}

\begin{figure*}
	\centering
	\begin{subfigure}{.51\textwidth}
		\centering
		\includegraphics[scale=0.41]{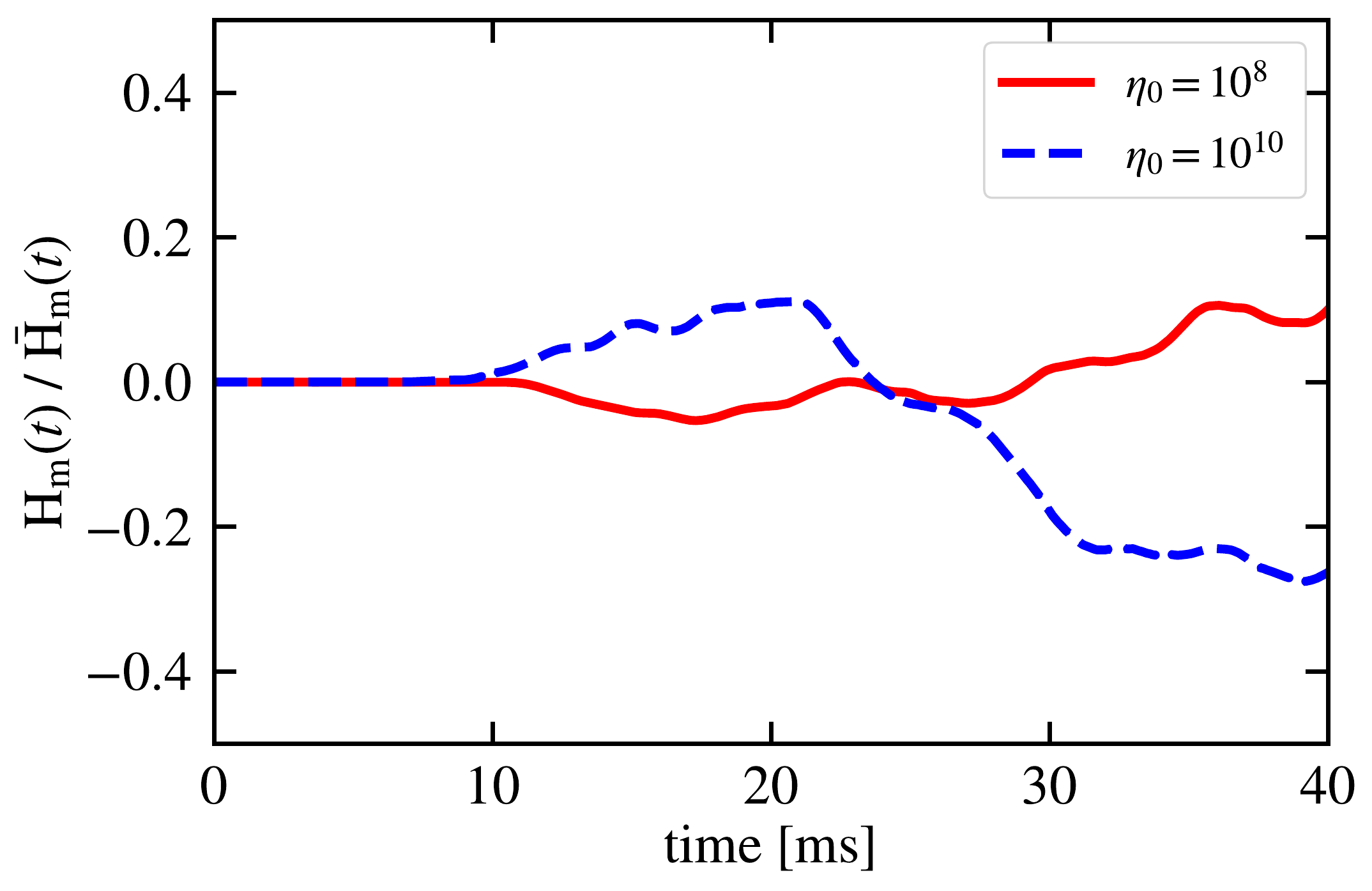}
		\caption{}
		\label{helicityplot2}
	\end{subfigure}%
	\begin{subfigure}{.51\textwidth}
		\centering
		\includegraphics[scale=0.41]{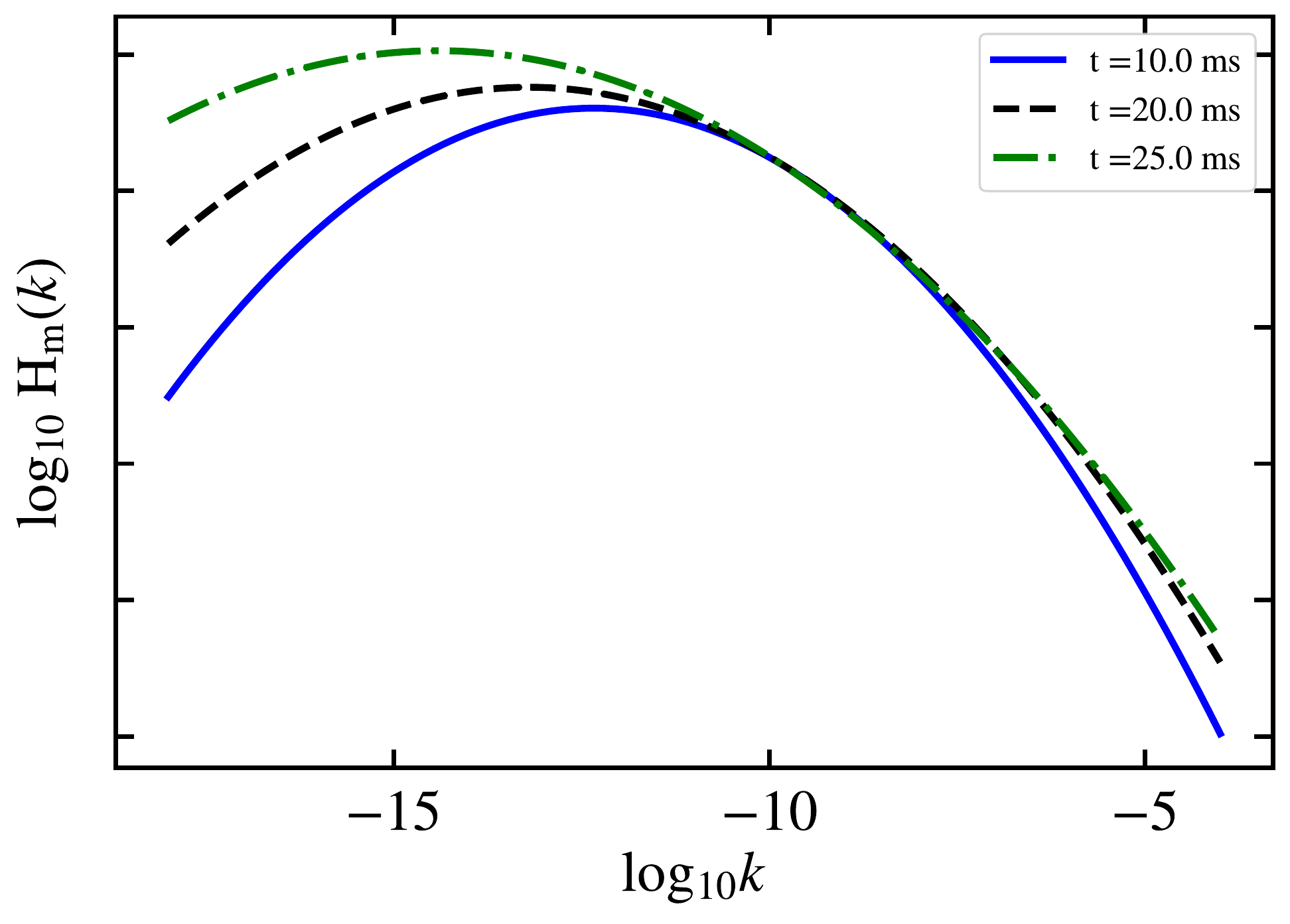}
		\caption{}
		\label{Maghelspectrum}
	\end{subfigure}
	\caption{(a) Magnetic helicity plotted as a function of time for two different values of $\eta_0$ for a setup with $\tau_{A} \sim 10$ s. The helicity becomes nonzero as the instability sets in after one {\alfven} crossing time. (b) The magnetic helicity spectrum at different times showing the phenomenon of an inverse cascade from the resistive small scales to larger scales.}
	
\end{figure*}

\vspace{0.4 cm}
\section{Turbulence \label{turbulence}}
\label{turbulence}
The presence of turbulence which drives the evolution and properties of the systems in the presence of an embedded magnetic field is very prominent in astrophysics, e.g. in accretion disks, interstellar medium, stellar winds, etc. 
In fact X-ray observations reveal that the magnetic field of the sun is in a turbulent state \citep{TurboSun}. We expect NSs to be turbulent soon after their birth where the heat and the escaping neutrinos provide the energy source. The turbulence decays after a short timescale ($\sim $ 1 day) as this energy source disappears, but is likely to play a role in the development of the field on the short timesales of hours we are examining, before the crust solidifies. Furthermore, even as the star evolves, differencse in angular velocity between the superfluid and the rest of the star are likely to lead to turbulence \citep{Turb1, Turb2}. Our simulations show that initial magnetic and kinetic energy drives the turbulence and the star reaches a turbulent `equilibrium', where average quantities can be studied, but in which the field is far from a stationary dipole. The non-linear saturation of the initial instabilities, lead, through the action of a small scale dynamo, to a turbulent mixed toroidal-poloidal field configuration, in which the ratio, averaged over the volume, of the energies in the two components reaches an equilibrium. When the Hall effect starts to dominate, this will also contribute to the development of turbulence \citep{Wareing09}.\\


To quantify this statement we start by studying the distribution of kinetic energy over different length scales. We plot the spectra for the kinetic and magnetic energies as a function of wavenumbers ($k$) in figure \ref{Kol}. As expected, the system shows higher dissipation of energy for smaller scale eddies through viscosity. Thus, the dynamics inside the star is turbulent. The classical Kolmogorov theory (\cite{Kol1941}) predicts that the turbulent energy spectrum in incompressible\footnote{Given the high speed of sound in NSs, we expect most eddies to be sub-sonic, and the effects of compressibility to be negligible.} hydrodynamic turbulence follows $E(k) \propto  k^{-5/3}$, where k is the modulus of the wave-vector ($k=\sqrt{k_x^2+k_y^2+k_z^2}$). In order to calculate the energy spectrum, we convert each velocity component into Cartesian space and Fourier transform them according to

\begin{eqnarray}
\bm{u}(\bm{k}) = \iiint_{\mathbb{R}^3} \bm{\hat{u(\bm{x})}}e^{i\bm{k}.\bm{x}}d^3\bm{x} 
\end{eqnarray}
\\
In figure \ref{Kolres}, we plot the Kolmogorov spectrum for kinetic energy for our resistive setup. 
Our spectra are consistent with a Kolmogorov like dependence, however in MHD turbulence, where the main interaction happens within wave-packets moving with {\alfven} velocities, the scaling relation follows $E(k) \propto k^{-3/2}$ (\cite{Iroshnikov1964, Kraichnan1965}). While previous numerical work has revealed the presence of a Kolmogorov spectrum also in MHD turbulence \citep{biskamp2003}, consistently with our interpretation, it is not possible to exclude $k=-3/2$ and determine the exact scaling relation with our limited resolution. Future high resolution studies are required to determine the exact nature of the turbulence in the NS problem. We note that the magnetic Reynolds number in MHD-turbulence is $R_m = {Lv/ \eta_0} = 10^6$, and since, $R_m >> 1$, the magnetic field lines are advected with the fluid flow and diffusion is unimportant.

\begin{figure*}
	\centering
	\begin{subfigure}{.55\textwidth}
		\centering
		\includegraphics[scale=0.461]{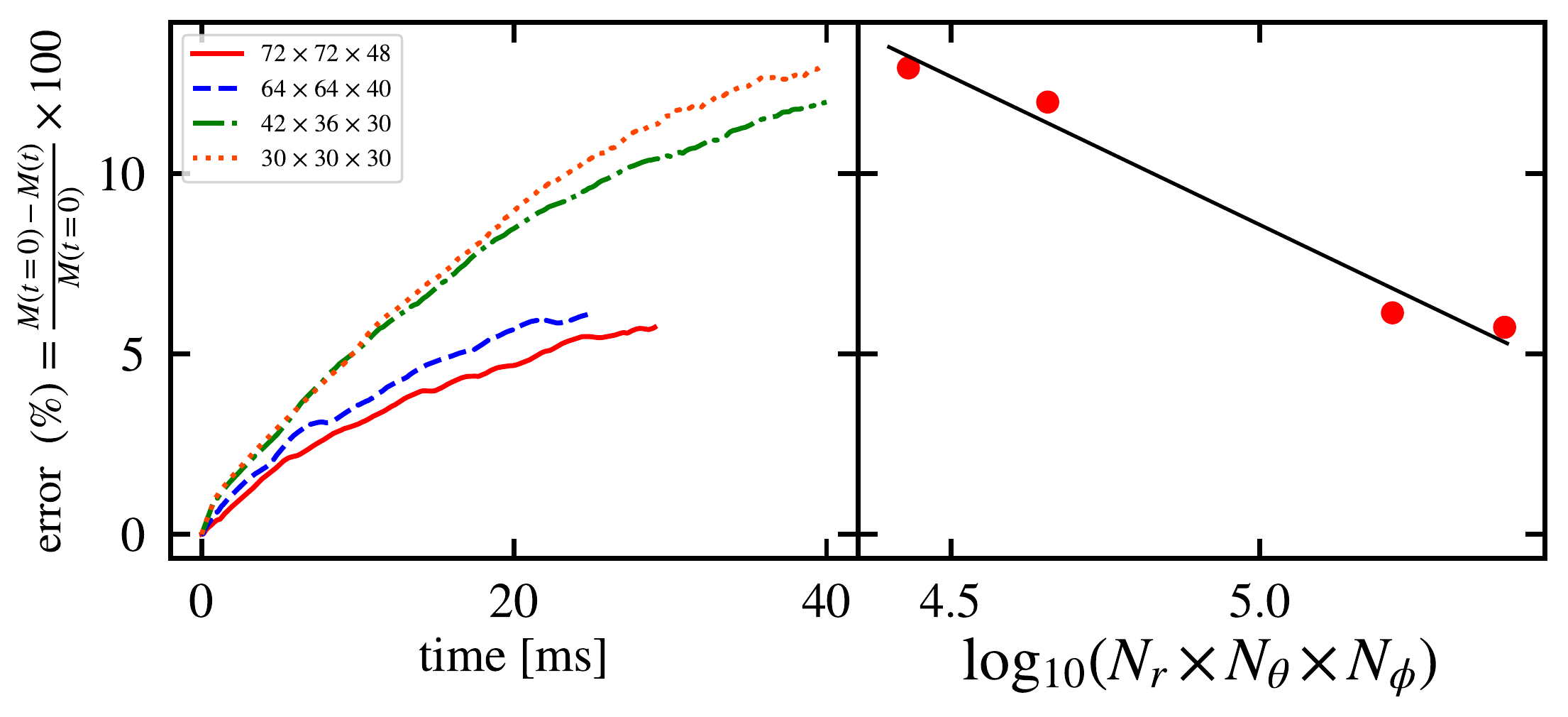}
		\caption{}
		\label{masserror}
	\end{subfigure}%
	\begin{subfigure}{.5\textwidth}
		\centering
		\includegraphics[scale=0.34]{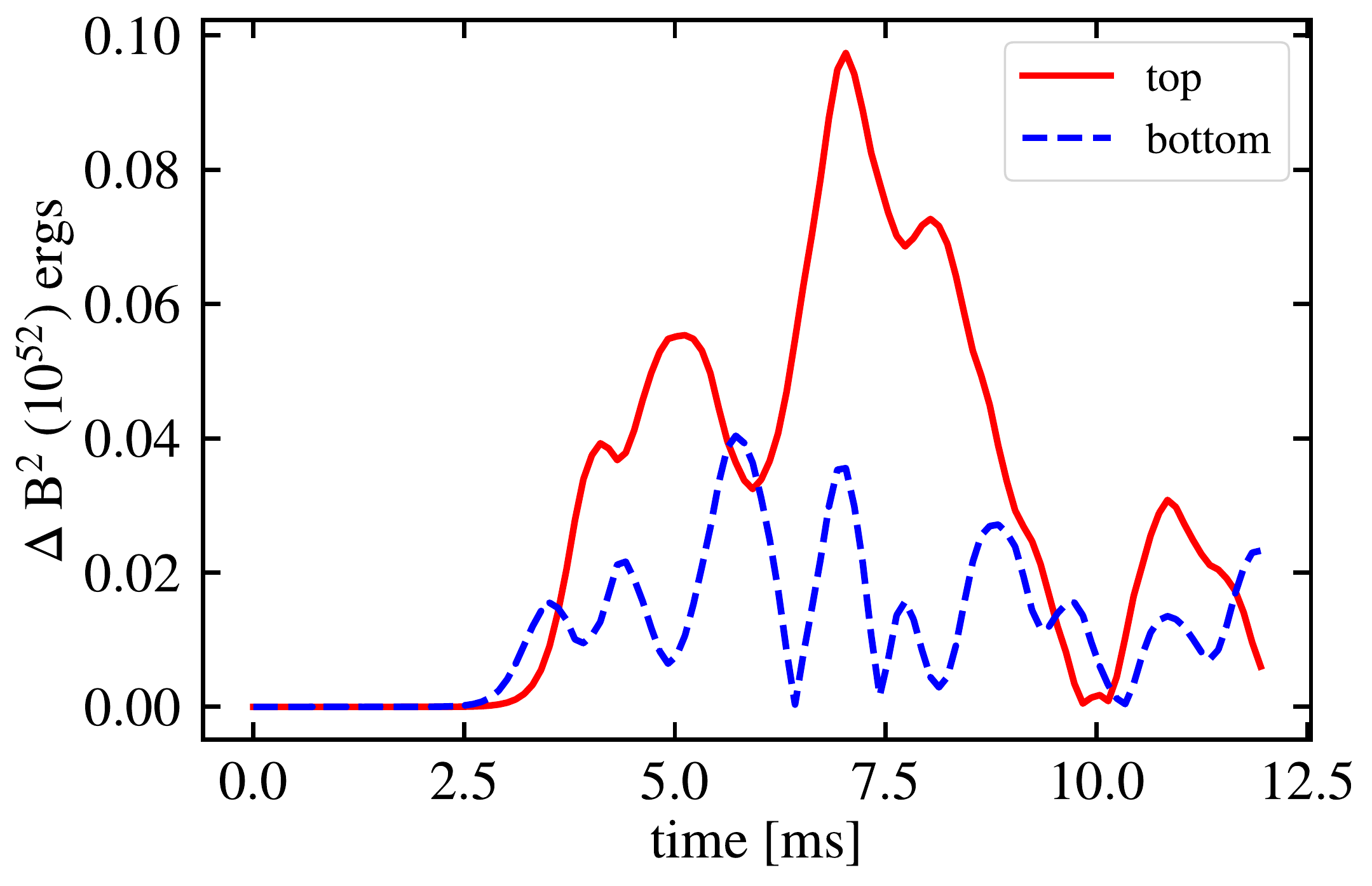}
		\caption{}
		\label{topbottom}
	\end{subfigure}
	\caption{(a) Error in the total mass of the star as a function of (left) time, and (right) resolution for our purely poloidal setup with $B_p = 10^{17}$ G. The curves are incomplete for higher resolution setups as the simulation hit the wall clock time. (b) Difference in $B_{\phi}^2$ plotted as a function of time. The red curve (top) shows the difference in energies between our setups with resolution $64\times64\times40$ and $30\times30\times30$. The blue curve (bottom) shows the difference in energies between our setups with resolution $72\times72\times48$ and $64\times64\times40$.}
	
\end{figure*}

\subsection{Cascade directions}

As discussed before, the magnetic helicity ($\rm H_{\rm m}$) is an ideal MHD invariant and as such its spectral density is conserved in nonlinear interactions. Figure \ref{helicityplot2} shows the variation of $\rm H_m$ with time for two different values of $\eta_0$. However, as turbulence is excited by the magnetic instabilities, $\rm H_{\rm m}$ is created on the resistive scale on which non-ideal effects (dictated in our case by numerical resistivity) act, i.e. its value becomes non-zero, and it is scattered to different length scales. This transfer proceeds from larger to smaller wavenumbers showing an \textit{inverse cascade} \citep{Frisch1975}, as the system attempts to conserve $\rm H_{\rm m}$ by moving it to scales much larger than the resistive scale. With increasing time, the peak of the magnetic helicity spectrum shifts to smaller $k$ showing the inverse cascade phenomenon, as seen in figure \ref{Maghelspectrum}.

This confirms the picture that turbulence plays a key role, by allowing to generate helicity in the system at small scales, and transfer it from the turbulent small scale structures to the larger scale magnetic field, thus creating a twisted-torus structure.

We note that at the end of our simulations turbulence has not decayed, but the average quantities, such as the average field strengths and magnetic energies have reached an equilibrium which is roughly constant over many $\alfven$ timescales. Longer simulations are needed to study the decay of turbulence, and understand whether in this case additional instabilities will appear also in our barotropic setup, as suggested by \citet{Mitchell2015}.


\section{Convergence \label{sec:convergence}}
\begin{figure*}
	\centering
	\begin{subfigure}{.51\textwidth}
		\centering
		\includegraphics[scale=0.4]{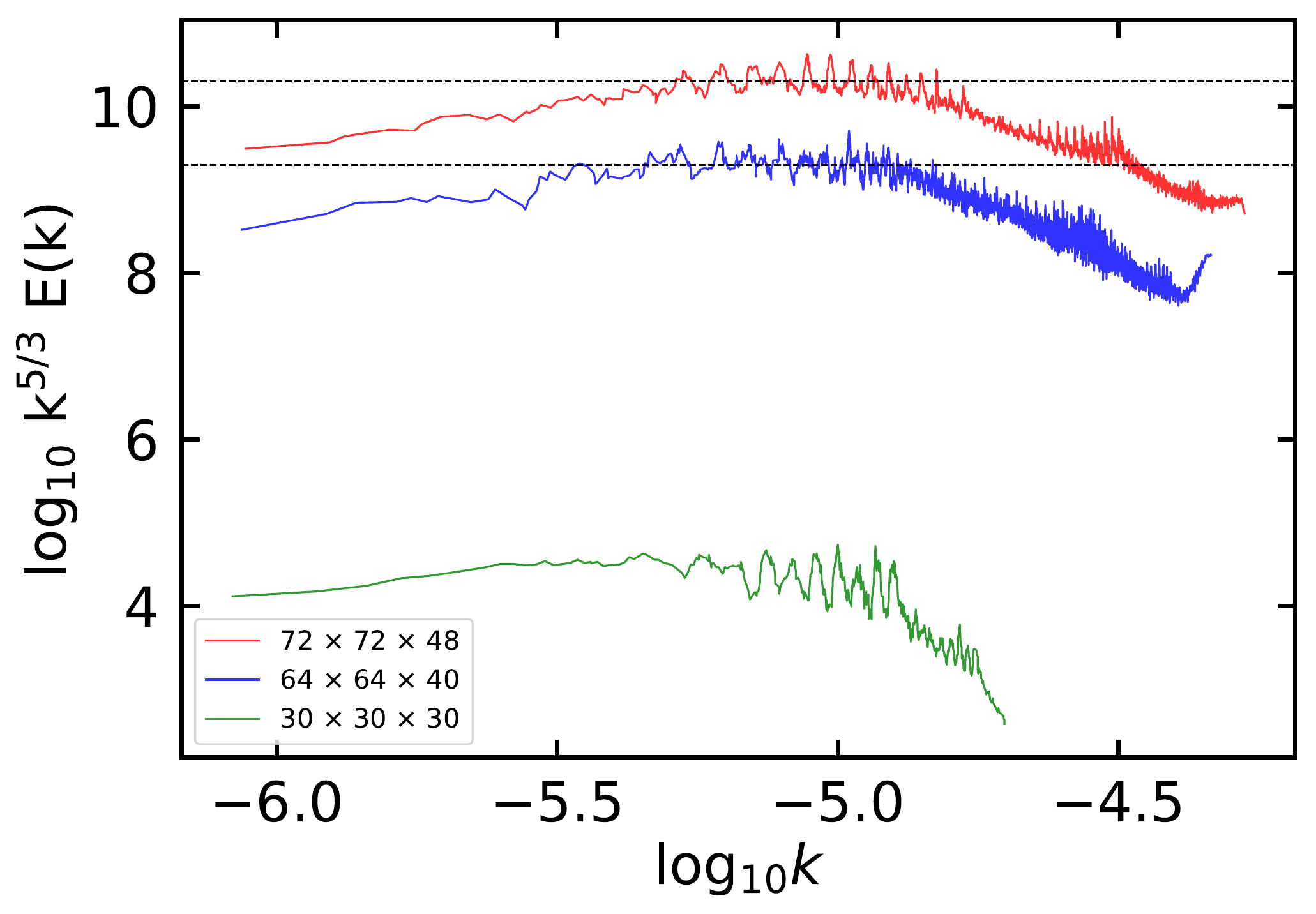}
		\caption{}
		\label{Kolmultiple}
	\end{subfigure}%
	\begin{subfigure}{.51\textwidth}
		\centering
		\includegraphics[scale=0.41]{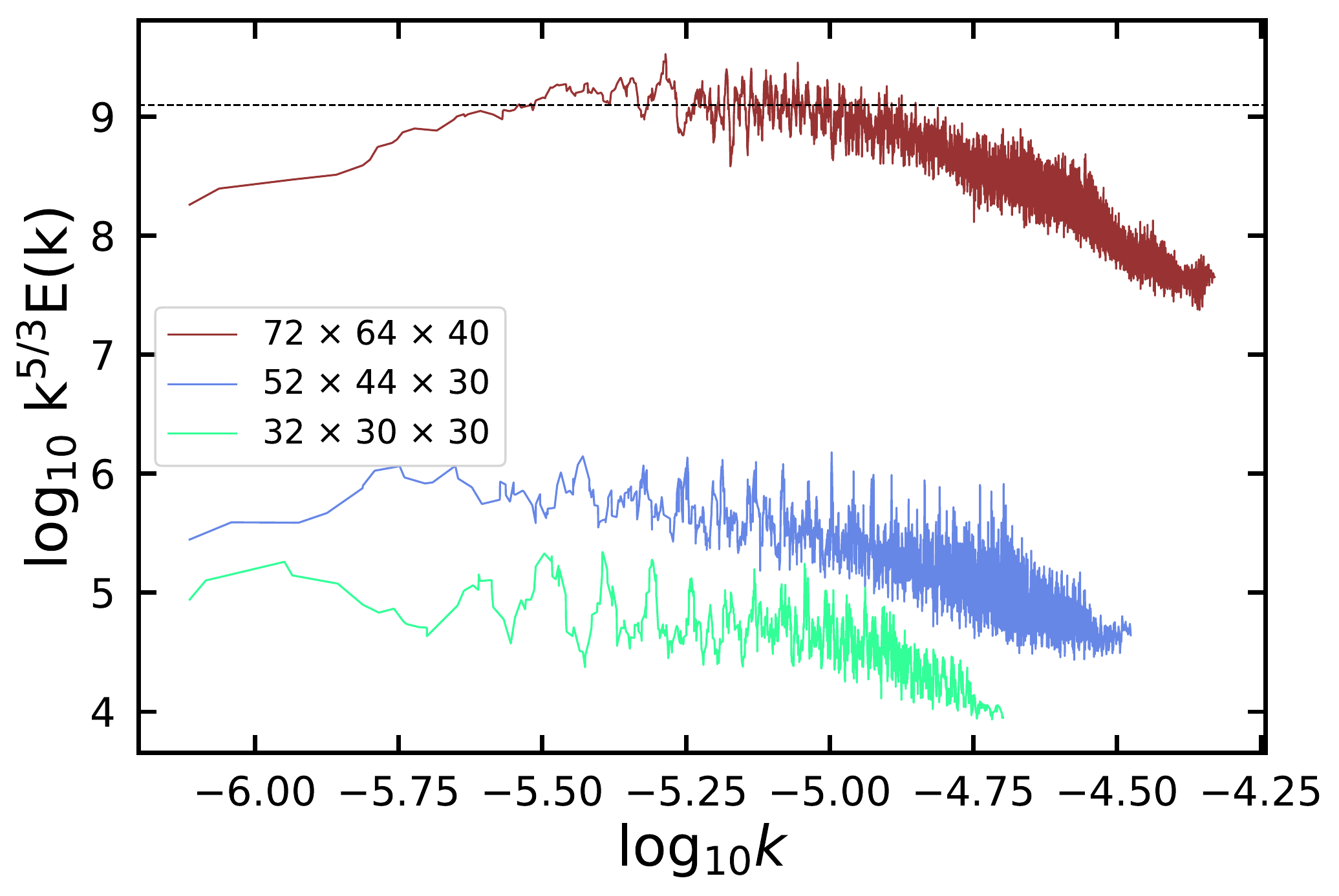}
		\caption{}
		\label{Kolresmultiple}
	\end{subfigure}
	\caption{(a) Kolmogorov spectrum plotted for our non-resistive setup with varying resolution. According to Classical Kolmogorov theory, the spectrum follows $E(k) \propto k^{-5/3}$ in the inertial range, and deviate for high and low values of $k$ where energy is injected and dissipated. We multiply $E(k)$ by $k^{5/3}$ and therefore expect to obtain a flat spectrum in the inertial range for the Kolmogorov case. This is seen in the region between $k = (10^{-5.25},10^{-4.9})$ for our highest resolution simulation (the black dotted lines are for reference). (b) Kolmogorov spectrum plotted for different resolutions for resistive atmosphere setup. Although the spectra are noisy, a Kolmogorov like dependence is visible in both the setups.}
	
\end{figure*}

In this section, we present our convergence tests focusing in particular on the non-resistive setup. Figure \ref{masserror} shows the variation in error of the total mass of the star with time and number of points in our grid respectively. 

The mass of the star as a function of time was calculated assuming spherical symmetry. However, the poloidal field makes the star oblate and pushes material out which causes the loss of spherical symmetry. This effect is not taken into account in our calculation, as it is expected to deform the spherical profile of the star by less than $\approx 0.01\%$ for a field of $B=10^{16}$ G \citep{Haskell:2007bh}. In fact, we see that the highest resolution has the least error in the total mass. We find that the error reduces with an increase in the number of grid-points, and the mass determination appears to converge.

We also analyse the energy in the toroidal field, which is one of our key observables. In this case point-wise convergence is almost certainly lost, as turbulence develops. If the code is converging, we expect the difference in energies for our middle-lowest resolution setup (defined as ``top'') to be higher than the difference in energies for the highest-middle resolution (defined as ``bottom''). This is illustrated in figure \ref{topbottom} where $B = B_{\phi}$. Although, the plots are oscillating, the expected trend is seen, and at later times, when turbulence is fully developed, convergence is worse and at times lost.

As turbulence affects the dynamics of the field, and affects the convergence of our results, we use the spectrum of the turbulence itself as a diagnostic for convergence. We have already analysed the spectrum for our higher resolution simulations in the previous sections, and found it to be consistent with a Kolmogorov spectrum. In figures (\ref{Kolmultiple}) and (\ref{Kolresmultiple}) we plot the kinetic energy spectrum for varying resolutions as a sanity check of the convergence of the code. As can be seen, the spectrum extends to smaller scales as expected, and is consistent with the scaling of 
\begin{eqnarray}
E(k) \propto  k^{-5/3}
\end{eqnarray}
over a larger portion of parameter space, indicating that our higher resolution simulations are increasingly capturing the true dynamics of the system.

\vspace{0.4 cm}
\section{Conclusions \& Discussions \label{sec:discussion}}
In this paper we have presented the results of three dimensional MHD simulations of magnetic field configurations in NSs. We have considered both ideal MHD and a setup with a resistive atmosphere, and assume the field to be dipolar at the exterior boundary far from the star. We do not consider the effect of the crust, or of superfluidity in the interior. Our results are thus applicable to the first few hours of life of the star, after differential rotation is dissipated. The field configurations we obtain are then 'frozen in' as the star cools, and may be sued as initial conditions for longer term simulations, on timescales of $10^3-10^5$ years, where the evolution of the field is driven by effects such as the Hall effect in the crust, Ohmic decay and ambipolar diffusion.

We have studied the evolution of both initially purely poloidal and mixed poloidal-toroidal fields with stronger toroidal components, and find that in all cases the initial configuration is unstable, with the instability developing on the order of an $\alfven$ crossing time scale. As the instability develops it gives rise to turbulence, and drives a small scale dynamo, which transfers helicity to the large scale field. The field attends a complex geometry with the toroidal component contributing $E_{\rm tor} \leq 20 \, \%$ of $E_{\rm m, tot}$ in all setups, and while this is not a strict equilibrium, the ratio of the poloidal to toroidal energies in the field is approximately stable.
The turbulence is not observed to decay during our simulations.

We find that stronger resistivity triggers the instability faster, but does not impact its non-linear saturation, thus modifying only the timescales in our simulation. We also found that the extent of the atmosphere does not play any role in the overall equilibrium of the system, and the results do not change if we push the boundary of our simulation farther out, from 1.2 to 2 stellar radii. Our results show that the field doesn't decay unlike the works of \cite{Braithwaite:2005md} and \cite{Mitchell2015}. Our choice of fixed boundary conditions could play a major role here and thus future studies will be aimed at understanding this scenario better.

Overall we find that a NS with a given inferred dipolar field strength far from the surface, is likely to harbour an interior toroidal component  with an average energy of roughly $25\%$ of the poloidal component, but that stronger toroidal fields are unstable and cannot be sustained. The overall geometry of the field is however complex, with higher multipoles growing closer to the surface, and more-over non-stationary over the life time of our simulations. We find rather a turbulent quasi-equilibrium, in which only average quantities are roughly constant.
Further studies will focus on the decay of the turbulence and on quantifying the impact of these results on attempts to measure the mass and radius of a NS with X-ray observations from NICER, for which the background field configuration is an important ingredient \citep{Nicer1, Nicer2, Nicer3}.

\section{Acknowledgement}
We would like to thank Samuel Lander, Paul Lasky, Andreas Reisenegger and Filippo Anzuini for useful comments and suggestions. AS gratefully acknowledges the help of Miljenko Cemeljic, Dipanjan Mukherjee and Varadaranjan Parthsarathy for teaching PLUTO. AS further thanks Marco Antonelli for helping with the Kolmogorov spectrum. BH thanks Sebastiano Bernuzzi and David Hilditch for useful comments. This project was supported by an OPUS grant from the National Science Centre, Poland (NCN), number 2018/29/B/ST9/02013. 

\bibliographystyle{mnras}
\bibliography{references}

\end{document}